\documentclass[12pt]{JHEP3}
\usepackage{epsfig            }
\usepackage{axodraw}

\epsfclipon
\usepackage{multicol}
\usepackage{epsfig,bm}
\usepackage{amssymb,amsmath}
\usepackage{graphicx}

\newcommand{\be}{\begin{equation}}
\newcommand{\ee}{\end{equation}}
\newcommand{\bea}{\begin{eqnarray}}
\newcommand{\beas}{\begin{eqnarray*}}
\newcommand{\eea}{\end{eqnarray}}
\newcommand{\eeas}{\end{eqnarray*}}
\newcommand{\ba}{\begin{array}}
\newcommand{\ea}{\end{array}}
\def\ls{\mathrel{\lower4pt\vbox{\lineskip=0pt\baselineskip=0pt
           \hbox{$<$}\hbox{$\sim$}}}}
\def\gs{\mathrel{\lower4pt\vbox{\lineskip=0pt\baselineskip=0pt
           \hbox{$>$}\hbox{$\sim$}}}}
\def\smiley{\hbox{\large$\bigcirc$\hspace{-.80em}%
\raise.2ex\hbox{$\cdot\cdot$}\kern-.61em    %--- .56
\lower.2ex\hbox{\scriptsize$\smile$}}\ }

\newcommand{\roughly}[1]{\mathrel{\raise.3ex\hbox{$#1$\kern-0.85em
\lower1ex\hbox{$\sim$}}}}

\newcommand{\lsim}{\roughly<}
\newcommand{\gsim}{\roughly>}

\def\be{\begin{equation}}
\def\beq\begin{equation}
\def\ee{\end{equation}}
\def\bea{\begin{eqnarray}}
\def\eea{\end{eqnarray}}

\def\beq{\begin{equation}}
\def\eeq{\end{equation}}
\def\beqa{\begin{eqnarray}}
\def\eeqa{\end{eqnarray}}

\newcommand{\bmat}{\left(\begin{array}}
\newcommand{\emat}{\end{array}\right)}

\title{Supersymmetric Thermalization and Quasi-Thermal Universe:
Consequences for Gravitinos and Leptogenesis } 

\author{Rouzbeh Allahverdi$^{1}$ and Anupam 
Mazumdar$^{2}$\\

$^{1}$~Theory Group, TRIUMF, 4004 Wesbrook Mall, Vancouver, BC, 
V6T 2A3, Canada. \\ 
$^{2}$~NORDITA, Blegdamsvej-17, Copenhagen-2100, Denmark.}

\abstract{ Motivated by our earlier paper~\cite{am}, we discuss how
the infamous gravitino problem has a natural built in solution within
supersymmetry. Supersymmetry allows a large number of flat directions
made up of {\it gauge invariant} combinations of squarks and
sleptons. Out of many at least {\it one} generically obtains a large
vacuum expectation value during inflation.  Gauge bosons and Gauginos
then obtain large masses by virtue of the Higgs mechanism.  This makes
the rate of thermalization after the end of inflation very small and
as a result the Universe enters a {\it quasi-thermal phase} after the
inflaton has completely decayed. A full thermal equilibrium is
generically established much later on when the flat direction
expectation value has substantially decareased. This results in low
reheat temperatures, i.e., $T_{\rm R}\sim {\cal O}({\rm TeV})$, which
are compatible with the stringent bounds arising from the big bang
nucleosynthesis. There are two very important implications: the
production of gravitinos and generation of a baryonic asymmetry via
leptogenesis during the quasi-thermal phase. In both the cases the
abundances depend not only on an effective temperature of the
quasi-thermal phase ( which could be higher, i.e., $T\gg T_{\rm R}$),
but also on the state of equilibrium in the reheat plasma. We show
that there is no ``thermal gravitino problem'' at all within
supersymmetry and we stress on a need of a new paradigm based on a
``quasi-thermal leptogenesis'', because in the bulk of the parameter
space the {\it old} thermal leptogenesis cannot account for the
observed baryon asymmetry.}

\preprint{NORDITA-2005-82.}

\begin{document}

%%%%%%%%%%%%%%%%%%%%%%%%%%%%%%%%%%%%%%%%%%%%%%%%%%%%%%%%%%%%%%%%%%%
%%%%%%%%%%%%%%%%%%%%%%%%%%%%%%%%%%%%%%%%%%%%%%%%%%%%%%%%%%%%%%%%%%%

\section{Introduction}

Primordial inflation~\cite{infl}~\footnote{For a realistic toy model
of inflation where the inflaton carries the charges of minimal
supersymmetric Standard Model, and gives rise to a low scale
inflation, see~\cite{MSSM-infl}.} is the best paradigm for explaining
the initial conditions for the structure formation and the cosmic
microwave background (CMB) anisotropies~\cite{wmap}. The Universe is
cold and empty after inflation and all the energy is stored in the
inflaton condensate oscillating around the minimum of its
potential. Reheating is an important stage for any inflationary
model. It describes the transition from this frozen state to a hot
thermal Universe. It involves various processes and eventually results
in a thermal bath of elementary particles which contains the Standard
Model (SM) degrees of freedom.

The Inflaton decay is the most relevant part of the reheating.  Only
one-particle decay of the non-relativistic inflaton quanta were
considered initially~\cite{reheat}~\footnote{For a recent discussion
on supersymmetric reheating, see~\cite{reheat-new}.}.  The treatment is
valid if the energy transfer to the fields which are coupled to the
inflaton takes place over many inflaton oscillations.  This requires
that the inflaton couplings to the SM fields are sufficiently small.
Usually it is assumed that the plasma reaches complete kinetic and
chemical equilibrium immediately after all the inflaton quanta have
decayed.  The reheat temperature of the Universe, $T_{\rm R}$, is then
quoted as:~\cite{reheat}
\begin{equation}
\label{reh1}
T_{\rm R}\sim 0.5 g_{\ast}^{-1/4}\sqrt{M_{P}\Gamma_{\rm d}}\,,
\end{equation}
where $g_{\ast}$ is the total number of relativistic degrees of
freedom, $\Gamma_{\rm d}$ is the inflaton decay rate and $M_{\rm P} = 2.4 
\times 10^{18}$ GeV is the reduced Planck mass.

It was also realized that the coherent oscillations of the inflaton
can create particles non-perturbatively~\cite{preheat1,preheat2}. This
mechanism is called preheating and it is particularly efficient when
the final products are bosonic degrees of freedom. It only takes about
two dozens of oscillations to transfer the energy from the homogeneous
condensate to non-zero modes of the final state(s)~\cite{preheat2}.
However, despite efficiently transferring the energy, preheating does
not result in a {\it complete} decay of the inflaton. An epoch of
perturbative reheating is an essential ingredient of any potentially
realistic cosmological model~\cite{jed}~\footnote{Preheating ends due
to backreaction as well as the expansion of the universe. Preheating
does not destroy the zero mode of the inflaton condensate completely,
although the amplitude of the inflaton oscillations diminish, but the
inflaton decay is completed when the zero mode perturbatively decays
into the SM or some other degrees of freedom,
see~\cite{preheat2}. Under only special conditions the inflaton
condensate can completely fragment,
see~\cite{kasuya}.}~\footnote{Besides, the initial stages of preheating
can give rise to a large non-Gaussianity which already rule out the
bulk of the parameter space from the current
observations~\cite{asko1}, see for a detailed discussion in our recent
paper~\cite{asko2}.}. In supersymmetry (SUSY) a first stage of
preheating is naturally followed by the last stage of perturbative
decay (see Appendix~\ref{perturbative}).  For these reasons we
concentrate on the perturbative inflaton decay, because of its
relevance to create a thermal bath of SM particles~\footnote{In
supergravity inflationary models the inflaton can be a hidden sector
which interacts gravitationally with other sectors. In which case the
inflaton decay is perturbative {\it at all times}. In string theory
however the epoch of thermalization could be very complicated as
described in the recent paper~\cite{Frey1}. This is due to the fact
that string thermodynamics differs in many respects to that of a
particle thermodynamics.}.

Often cosmology is considered as a probe to the early Universe, a well
known example is the {\it Gravitino Problem} in the context of a
supersymmetric cosmology. SUSY introduces new degrees of freedom and
new parameters. Most of them are rather poorly constrained from
experiments.  Cosmology however acts as a test bed where {\it some of
the SUSY particles} can be tested.  In this regard the reheat
temperature plays an important role as we shall explain below.

The gravitino is a spin-$3/2$ supersymmetric partner of the graviton,
which is coupled to the SM particles with a gravitational strength.
The Gravitinos with both the helicities can be produced from a thermal
bath. There are many scattering channels which include fermion,
sfermion, gauge and gaugino quanta all of which have a cross-section
$\propto 1/M^2_{\rm P}$~\cite{moroi}, which results in a gravitino
abundance (up to a logarithmic correction)
as~\cite{thermal,gmsb}~\footnote{From now on we take $g_{\ast} =
228.75$ as in the case of minimal supersymmetric SM (MSSM).}:
\begin{eqnarray} \label{gravtherm}
{\rm Helicity}~ \pm {3 \over 2}: {n_{3/2} \over s} 
&\simeq & \left({T_{\rm R} \over 10^{10}~{\rm GeV}}\right) ~ 
10^{-12}\,, \nonumber \\
& & \hspace{7cm} ({\rm full ~ equilibrium}) \, \nonumber \\
{\rm Helicity} ~ \pm {1 \over 2}: {n_{3/2} \over s}
&\simeq & \left(1 + {M^2_{\widetilde g} \over 12 m^2_{3/2}}\right) 
\left({T_{\rm R} \over 10^{10}~{\rm GeV}}\right) ~ 10^{-12}\,;
\end{eqnarray}
where $M_{\widetilde g}$ is the gluino mass. Note that for
$M_{\widetilde g} \leq m_{3/2}$ both the helicity states have
essentially the same abundance, while for $M_{\widetilde g} \gg
m_{3/2}$ production of helicity $\pm 1/2$ states is enhanced due to
their Goldstino nature. The linear dependence of the gravitino
abundance on $T_{\rm R}$ can be understood qualitatively. Since the
cross-section for the gravitino production is $\propto M^{-2}_{\rm
P}$, the production rate at a temperature, $T$, and the abundance of
the gravitinos produced within one Hubble time will be $\propto T^3$
and $\propto T$ respectively. This implies that the gravitino
production is efficient at the highest temperature of the
radiation-dominated phase of the Universe, i.e., $T_{\rm R}$.

An unstable gravitino decays to particle-sparticle pairs, and its
decay rate is given by $\Gamma_{3/2} \simeq m^{3}_{3/2}/4 M^2_{\rm
P}$~\cite{moroi}.
%
%\begin{equation} \label{decrate}
%\Gamma_{3/2} \simeq \left(N_g+\frac{N_{f}}{12}\right) ~ 
%\frac{m^3_{3/2}}{32 \pi M^2_{\rm P}}\,, 
%\end{equation}
%
%where $N_g$ and $N_f$ are the number of available decay channels into
%gauge-gaugino and fermion-sfermion pairs respectively. 
%The gravitino
%decay is completed when $H \simeq \Gamma_{3/2}$, when the
%temperature of the Universe is given by
%
%\begin{equation} \label{dectemp}
%T_{3/2} \simeq \left(\frac{m_{3/2}}{10^{5}~{\rm GeV}}\right)^{3/2}~7~
%{\rm MeV}\,.
%\end{equation}
%
If $m_{3/2} < 50$ TeV, the gravitinos decay during or after big bang
nucleosynthesis (BBN)~\cite{bbn}, which can ruin its successful
predictions for the primordial abundance of light
elements~\cite{subir}. If the gravitinos decay radiatively, the most
stringent bound, $\left(n_{3/2}/s\right) \leq 10^{-14}-10^{-12}$,
arises for $m_{3/2} \simeq 100~{\rm GeV}-1$ TeV~\cite{cefo}.

On the other hand, much stronger bounds are derived if the gravitinos
mainly decay through the hadronic modes. In particular, for a hadronic
branching ratio $\simeq 1$, and in the same mass range,
$\left(n_{3/2}/s\right) \leq 10^{-16}-10^{-15}$ will be
required~\cite{kkm}.

For a radiatively decaying gravitino the tightest bound
$\left(n_{3/2}/s \right) \leq 10^{-14}$ arises when $m_{3/2} \simeq
100$ GeV~\cite{cefo}.  Following Eq.~(\ref{gravtherm}) the bound on
reheat temperature becomes: $T_{\rm R} \leq 10^{10}$ GeV. This turns
out to be a very stringent limit if the inflaton decay products
immediately thermalize.  In fact, for the inflaton mass $m_{\phi} =
10^{13}$ GeV, it is at best marginally satisfied even for a
gravitationally decaying inflaton (see an example given in an
Appendix~\ref{hidden}).  For a TeV gravitino which mainly decays into
gluon-gluino pairs (allowed when $m_{3/2} > M_{\widetilde g}$) a much
tighter bound $\left(n_{3/2}/s \right) \leq 10^{-16}$ is
obtained~\cite{kkm}, which requires quite a low reheat temperature:
$T_{\rm R} \leq 10^6$~GeV.

The gravitino will be stable if it is the lightest supersymmetric
particle (LSP), where $R$-parity is conserved. The gravitino abundance
will in this case be constrained by the dark matter
limit,~\cite{wmap}, $\Omega_{3/2} h^2 \leq 0.129$, leading to
\begin{equation} \label{dmlimit}
{n_{3/2} \over s} \leq 
5 \times 10^{-10} ~ \left({1~{\rm GeV} \over m_{3/2}}\right)\,.
\end{equation}
For $m_{3/2} < M_{\widetilde g}$, the helicity $\pm 1/2$ states
dominate the total gravitino abundance.  As an example, consider the
case with a light gravitino, $m_{3/2} = 100$~KeV, which can arise very
naturally in gauge-mediated models~\cite{dmm}.  If $M_{\widetilde g}
\simeq 500$ GeV, see Eq.~(\ref{gravtherm}), a very severe constraint,
$T_{\rm R} \leq 10^4$ GeV, will be obtained on the reheat
temperature~\footnote{Similar results were obtained in connection to
the brane worlds, see for instance~\cite{anup1}, and for the
discussions on reheating, see~\cite{anup2} and \cite{anup3}.}.

As we have argued, gravitinos indirectly put constraints on the
thermal history of the Universe, which is often known as the {\it
gravitino problem} in models with weak scale SUSY. The reheat
temperature, $T_{\rm R}$, plays a very important role here as it sets
the thermal (and possibly the largest) contribution to the gravitino
abundance. However besides the gravitino production, it has other
implications for cosmology, for instance, thermal leptogenesis,
thermal production of weakly interacting dark matter particles,
etc. They are all directly or indirectly connected to the reheat
temperature of the Universe.

The most important worry is the central assumption that, {\bf the SM
and or MSSM degrees of freedom thermalize instantly}. In our opinion
this is the {\it key assumption} which has not been questioned well
enough in the literature. It is only very recently that various issues
of thermalization have been considered in a NON-SUSY case
carefully~\cite{ds}. We intend to cover this lapse and describe
thermalization process within SUSY. The subject stands on its own
right because SUSY introduces new ingredients into the game, i.e.,
{\it flat directions}. We will elaborate on the role of flat
directions in thermalization, which was first pointed out in our
earlier paper~\cite{am}.

SUSY along with gauge symmetry introduces gauge invariant flat
directions. In any supersymmetric extension of the SM there are flat
directions primarily made up of squarks, sleptons (SUSY partners of SM
quarks and leptons) and Higgses. A flat direction in a cosmological
context can take a large vacuum expectation value (VEV) by virtue of a
shift symmetry. During inflation quantum fluctuations of any light
field accumulate in a coherent state and its VEV makes a random walk
with its variance growing linearly in time. The condensate becomes
homogeneous on scales larger than the size of the Hubble radius with a
growing VEV. However in many cases (e.g., minimal supergravity) shift
symmetry is not protected which truncates the VEV to be as large as
the four dimensional Planck scale, $M_{\rm P}$. Moreover SUSY is
broken and the flat directions obtain soft mass terms and
$A$-terms. It is also possible that non-renormalizable terms arise in
the superpotential after integrating out heavy degrees of freedom
which are associated to a new physics at high scales. These
contributions lift the flatness but still allow the VEV to be
significantly large in a wide class of models, for a review
see~\cite{flat}.

After inflation the VEVs of flat directions do not settle at the
origin immediately. The cosmologically sliding VEV plays a crucial
role in slowing down thermalization of the inflaton decay products as
mentioned in Ref.~\cite{am}. The idea is very simple, if a flat
direction develops a VEV, the SM gauge fields become massive, similar
to the Higgs mechanism. The masses for the gauge bosons (and gauginos)
are proportional to the VEV of the flat direction.  This obviously
breaks the charge and the color in the early Universe, but the VEV of
the flat direction finally vanishes and therefore there is no threat
to the low energy phenomenology.

Consider a situation where the inflaton has completely decayed into
the MSSM degrees of freedom~\footnote{The flat directions could also
reheat the Universe and could also explain the fluctuations present in
the cosmic microwave background radiation,
see~\cite{flatdens,recent0}. However there is a distinction, in this
paper we demand that the inflaton decay is the major source for the
entropy production.}. The resulting plasma is initially far away from
a full thermal equilibrium.  The leading reactions which establish
equilibrium are the $2 \rightarrow 2$ and $2 \rightarrow 3$
scatterings mediated by gauge fields~\cite{ds,ad2,jm}.  The former
lead to kinetic equilibrium, while the latter are required to change
the total number of particles in order to give rise to a chemical
equilibrium.  However these processes are suppressed by the VEV
dependent masses of the gauge bosons (and gauginos).  As a result the
Universe enters a period of a {\it quasi-thermal} phase during which
the reheat plasma evolves adiabatically. This lasts until the above
mentioned scatterings become efficient, at which time full equilibrium
is finally attained, i.e., within one Hubble time.

Such a late thermalization also results in a low reheat
temperature. For the intermediate values of the flat direction's VEV,
the reheat temperature within MSSM varies:
\be
10^{3}~{\rm GeV}\leq T_{\rm R}\leq 10^{7}~{\rm GeV}\,.
\ee
Interestingly such a low reheat temperature does not lead to a
significant generation of the gravitinos~\footnote{Note that gravitinos
are still produced during
preheating~\cite{non-pert1,non-pert2,non-pert3} and the final stage of
perturbative decay of the inflaton~\cite{nop,ajm} (see
also~\cite{aem,ad4}). We will consider these cases separately in a
different publication.}.  On the other hand, this is a {\it bad news}
for thermal leptogenesis which typically requires that $T_{\rm R} \geq
10^9$~GeV~\cite{buchmuller}.

One of the aims of this paper is to explore an alternative
prescription for leptogenesis, here, we will introduce a new paradigm,
quasi-thermal leptogenesis, which points towards a successful
baryogenesis within SUSY.

The rest of this paper is organized as follows. In Section 2 we
briefly review thermalization after the perturbative inflaton decay
(including relevant processes which lead to kinetic and chemical
equilibrium), and then discuss the impact of supersymmetric flat
directions on thermalization. Through numerical examples, in Section
3, we will underline the dramatically altered picture which
emerges. In Section 4 we will consider thermalization in some of the
other supersymmetric extensions of the SM, i.e., models with gauge
mediation and split supersymmetry. After discussing particle
production in the quasi-thermal phase in Section 5, we will specialize
to the gravitino production and the leptogenesis in Sections 6 and 7
respectively. In Section 8 we will make some remarks on thermalization
after preheating. We have also included Appendix which would help the
paper to be self contained.

%%%%%%%%%%%%%%%%%%%%%%%%%%%%%%%%%%%%%%%%%%%%%%%%%%%%%
%%%%%%%%%%%%%%%%%%%%%%%%%%%%%%%%%%%%%%%%%%%%%%%%%%%%%
\section{Thermalization in supersymmetric theories: Part-1}
\label{TST-1}
\subsection{A brief recourse to thermalization}

For a plasma which is in full thermal equilibrium, the energy density,
$\rho$, and the number density, $n$, of a relativistic particles are
given by
\begin{eqnarray} \label{full}
\rho &=& \left({\pi^2/30}\right) T^4\,, ~~~~~~~~~~n = 
\left({\zeta(3)/\pi^2}\right)T^3\,, ~~~~ ({\rm Boson}) \, , 
\nonumber \\ 
\rho &=& \left({7/8}\right) \left({\pi^2/30}\right) T^4\,, ~~~ 
n = \left({3/4}\right) \left({\zeta(3)/\pi^2}\right) T^3\,, ~~~~ 
({\rm Fermion}) \, ,
\end{eqnarray}
where $T$ is the temperature of a thermal bath. Note that in a full
equilibrium the relationships, $\langle E \rangle \sim \rho^{1/4}$,
and $n \sim \rho^{3/4}$ hold, with $\langle E \rangle = \left(\rho/n
\right) \simeq 3 T$ being the average particle energy. On the other
hand, right after the inflaton decay has completed, the energy density
of the Universe is given by: $\rho \approx 3 \left(\Gamma_{\rm d}
M_{\rm P}\right)^2$. For a perturbative decay, which generates
entropy, (see Eq.~(\ref{finpert}) of Appendix~\ref{perturbative}), we
have $\langle E \rangle \approx m_{\phi} \gg \rho^{1/4}$. Then, from
the conservation of energy, the total number density is found to be,
$n \approx \left(\rho/m_{\phi}\right) \ll \rho^{3/4}$. Hence the {\it
complete inflaton decay} results in a dilute plasma which contains a
small number of very energetic particles. This implies that the
Universe is far from full thermal equilibrium initially.

Reaching full equilibrium requires re-distribution of the energy among
different particles, {\it kinetic equilibrium}, as well as increasing
the total number of particles, {\it chemical equilibrium}. Therefore
both the number-conserving and the number-violating reactions must be
involved.  Full equilibrium is achieved shortly after the
number-violating processes become efficient.  From then on, we have
the familiar hot big bang Universe, which consists of elementary
particles in full thermal equilibrium. The maximum temperature of the
Universe after {\it complete} thermalization is the definition of the
reheat temperature, i.e., $T_{\rm R}$.

Let us first begin with the number-conserving reactions which build
kinetic equilibrium among the SM fermions. The most important
processes are $2 \rightarrow 2$ scatterings with gauge boson exchange
in the $t$-channel, shown in Fig.~(1). The cross-section for these
scatterings is $\sim \alpha \vert t \vert^{-1}$.  Here $''t''$ is
related to the exchanged energy, $\Delta E$, and the momentum,
$\overset {\longrightarrow}{\Delta p}$, through $t = {\Delta E}^2 -
{\vert \overset {\longrightarrow}{\Delta p} \vert}^2$. The fine
structure constant is denoted by $\alpha$ (note that $\alpha \geq
10^{-2}$ in the MSSM).  This cross section can be understood as
follows: the gauge boson propagator introduces a factor of ${\vert t
\vert}^{-2}$, while phase space integration results in an extra factor
of $\vert t \vert$. Due to an infrared singularity, these scatterings
are very efficient even in a dilute plasma~\footnote{There are also $2
\rightarrow 2$ scattering diagrams with a fermion or scalar exchange
in the $t$-channel. Diagrams with a fermion (for example, gaugino)
exchange have an amplitude $\propto {\vert t \vert}^{-1}$, which will
be canceled by the phase space factor $\vert t \vert$. This results
in a much smaller cross-section $\propto s^{-1}$, where $s \approx 4
E^2$ is the squared of the center-of-mass energy. Diagrams with a
scalar (for example, Higgs) exchange are also suppressed by the
following reason. A fermion-fermion-scalar vertex, which arises from a
Yukawa coupling, flips the chirality of the scattered fermion. For
relativistic fermions, as we consider here, the mass is $\ll E$ and a
flip of chirality also implies a flip of helicity. This is forbidden
by the conservation of angular momentum for forward scatterings, i.e.,
where $t \rightarrow 0$. As a result, the diagram in Fig.~(2) has no
$t$-channel singularity at all. Note that it is additionally
suppressed by powers of Yukawa couplings compared to the diagram with
gauge interactions in Fig.~(1). For more details, see
Ref.~\cite{ad2}}.

Note that every degree of freedom of the MSSM has some gauge
interactions with all other fields. Therefore any fermion in the
plasma has $t-$channel scatterings off those fermions with the largest
number density. For this reason the total number density, $n$, enters
while estimating the scattering rate.

\vspace*{6mm}
\begin{center}
\SetScale{0.6} \SetOffset(50,40)
\begin{picture}(350,100)(0,0)
\ArrowLine(75,90)(175,90) \Text(30,65)[l]{$f_1$}
\Vertex(175,90){3}
\ArrowLine(75,10)(175,10) \Text(30,0)[l]{$f_2$}
\Gluon(175,90)(175,10){5}{4} 
\ArrowLine(175,90)(275,90) \Text(175,65)[r]{$f_1$}
\Vertex(175,10){3}
\ArrowLine(175,10)(275,10) \Text(175,0)[r]{$f_2$}
\end{picture}
\vspace*{-13mm}

\end{center}

\vspace*{-0mm}
\noindent
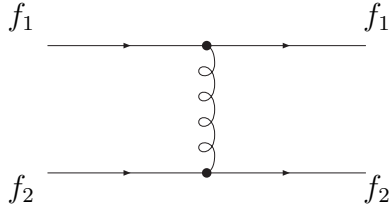
{\bf Fig. 1:}~Typical scattering diagram which builds kinetic
equilibrium in the reheat plasma. Note that the $t-$channel
singularity which results in a cross-section $\propto \vert t
\vert^{-1}$.

\vspace*{6mm}

In addition one also needs to achieve chemical equilibrium by changing
the number of particles in the reheat plasma. The {\it relative}
chemical equilibrium among different degrees of freedom is built
through $2 \rightarrow 2$ annihilation processes, occurring through
$s-$channel diagrams. Hence they have a much smaller cross-section
$\sim \alpha s^{-1}$.  More importantly the total number of particles
in the plasma must also change.  It turns out from Eq.~(\ref{full})
that in order to reach full equilibrium, the total number of particles
must {\it increase} by a factor of: $n_{\rm eq}/n$, where $n \approx
\rho/m_{\phi}$ and the equilibrium value is: $n_{\rm eq} \sim
\rho^{3/4}$. This can be a very large number, for examples given in
the appendix, $n_{\rm eq}/n\sim {\cal O}(10^3)$. This requires that
the number-violating reactions such as decays and inelastic
scatterings must be efficient.

Decays (which have been considered in Ref.~\cite{a}) are helpful, but in 
general they cannot increase the number of particles to the required level. 
It was recognized in~\cite{ds}, see also~\cite{ad2,jm}, that the most
relevant processes are $2 \rightarrow 3$ scatterings with gauge-boson
exchange in the $t-$channel. Again the key issue is the infrared
singularity of such diagrams shown in Fig.~(2). The cross-section for
emitting a gauge boson, whose energy is ${\vert t \vert}^{1/2} \ll E$,
from the scattering of two fermions is $\sim \alpha^3 {\vert t
\vert}^{-1}$. When these inelastic scatterings become efficient, i.e.,
their rate exceeds the Hubble expansion rate, the number of particles
increases very rapidly~\cite{es}, because the produced gauge bosons
subsequently participate in similar $2 \rightarrow 3$ scatterings.

As a result the number of particles will reach its equilibrium value,
$n_{\rm eq}$, soon after the $2 \rightarrow 3$ scatterings become
efficient. At that point the scatterings and inverse scatterings occur
at the same rate and the number of particles will not increase
further.  Therefore full thermal equilibrium will be established
shortly after the $2 \rightarrow 3$ scatterings become efficient. For
this reason, to a very good approximation, one can use the rate for
inelastic scatterings as a thermalization rate of the Universe
$\Gamma_{\rm thr}$.

\vspace*{6mm}
\begin{center}
\SetScale{0.6} \SetOffset(50,40)
\begin{picture}(350,150)(0,0)
\ArrowLine(75,90)(175,90) \Text(30,65)[l]{$f_1$}
\Vertex(175,90){3}
\ArrowLine(75,10)(175,10) \Text(30,0)[l]{$f_2$}
\Gluon(175,90)(175,10){5}{4} 
\ArrowLine(175,90)(275,90) \Text(180,65)[r]{$f_1$}
\Vertex(175,10){3}
\Vertex(215,90){3}
\Gluon(215,90)(275,150){5}{4}
\ArrowLine(175,10)(275,10) \Text(180,0)[r]{$f_2$}
\end{picture}
\vspace*{-13mm}
\end{center}

\noindent
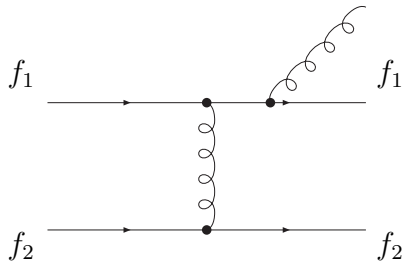
{\bf Fig. 2:}~Typical scattering diagrams which increase the number of 
particles.
\vspace*{2mm}

In order to estimate thermalization rate, we need to choose an
infrared cut-off on the parameter $t$. With a reasonable choice of a
cut-off, it turns out that in realistic models the $2 \rightarrow 3$
scatterings have a rate higher than or same as that of the inflaton
decay rate (for more details we refer the readers
to~\cite{ds,ad2,jm}).  Therefore, if the inflaton decay products have
gauge interactions, the Universe reaches full thermal equilibrium
immediately after the inflaton decay. The reason is that the $2
\rightarrow 3$ scatterings with gauge boson exchange in the
$t-$channel are very efficient.  However this analysis crucially
depends on having {\it massless gauge bosons}.

In general the reheat plasma induces a mass for all particles which it
contains. This is however negligible in a dilute plasma as in the case
after the perturbative inflaton decay.  However SUSY alters the
situation quite dramatically as we shall discuss below.

%%%%%%%%%%%%%%%%%%%%%%%%%%%%%%%%%%%%%%%%%%%%%%%%%%%%%%%%%%%%%%%%%%%%%%%%%%%%%
%%%%%%%%%%%%%%%%%%%%%%%%%%%%%%%%%%%%%%%%%%%%%%%%%%%%%%%%%%%%%%%%%%%%%%%%%%%%%
\subsection{Flat directions in Supersymmetry}
\label{FDS}

The field space of supersymmetric theories contains many directions
along which the $D-$ and $F-$term contributions to the scalar
potential identically vanish in the limit of unbroken SUSY. The most
interesting such directions are those made up of SUSY partners of the
SM fermions, namely the squarks and sleptons, and the Higgs
fields. These directions have gauge and Yukawa interactions with
matter fields, and hence decay before BBN.  Therefore they do not lead
to the cosmological moduli problem. As a matter of fact they can have
very interesting cosmological consequences (for a review, see
Ref.~\cite{flat}).

The superpotential for the Minimal Supersymmetric Standard Model
(MSSM) is given by, see for instance~\cite{nilles}
\begin{equation}
\label{mssm}
W_{MSSM}=\lambda_uQH_u u+\lambda_dQH_d d+\lambda_eLH_d e~
+\mu H_uH_d\,,
\end{equation}
where $H_{u}, H_{d}, Q, L, u, d, e$ in Eq.~(\ref{mssm}) are chiral
superfields representing the two Higgs fields (and their Higgsino
partners), LH (s)quark doublets, RH up- and down-type (s)quarks, LH
(s)lepton doublets and RH (s)leptons respectively.  The dimensionless
Yukawa couplings $\lambda_{u}, \lambda_{d}, \lambda_{e}$ are $3\times
3$ matrices in the flavor space, and we have omitted the gauge and
flavor indices. The last term is the $\mu$ term, which is a
supersymmetric version of the SM Higgs boson mass.

The SUSY scalar potential $V$ is the sum of the F- and D-terms and
reads
\begin{equation}
\label{fplusd}
V= \sum_i |F_i|^2+\frac 12 \sum_a g_a^2D^aD^a
\end{equation}
where
\begin{equation}
F_i\equiv {\partial W_{MSSM}\over \partial \chi_i},~~D^a=\chi^\ast_i T^a_{ij}
\chi_j~.
\label{fddefs}
\end{equation}
Here the scalar fields, denoted by $\chi$, transform under a gauge
group $G$ with the generators of the Lie algebra and gauge coupling
are by $T^{a}$ and $g_a$ respectively.

For a general supersymmetric model with $N$ chiral superfields
$X_{i}$, it is possible to find out the directions where the potential
in Eq.~(\ref{fplusd}) vanishes identically by solving simultaneously
\begin{equation}
\label{fflatdflat}
D^{a}\equiv X^{\dagger}T^{a}X=0\,, \quad \quad
F_{X_{i}}\equiv \frac{\partial W}{\partial X_{i}}=0\,.
\end{equation}
Field configurations obeying Eq.~(\ref{fflatdflat}) are called
respectively D-flat and F-flat.

D-flat directions are parameterized by gauge invariant monomials of
the chiral superfields. A powerful tool for finding the flat
directions has been developed in
\cite{buccella82,affleck84,ad,drt,luty96,gkm}, where the
correspondence between gauge invariance and flat directions has been
employed. There are nearly $300$ flat directions within
MSSM~\cite{gkm}.

The flat directions are lifted by soft SUSY breaking mass term,
$m_{0}$,
\begin{equation}
V\sim m_{0}^2|\varphi|^2\,,
\end{equation}
as well as higher order terms arising from the
superpotential~\cite{drt}~\footnote{Within supergravity there are
corrections to the SUSY potential arising from the K\"ahler potential
and also the mixing between superpotential and K\"ahler terms. We will
consider these issues in Section~\ref{part2}.}.

%%%%%%%%%%%%%%%%%%%%%%%%%%%%%%%%%%%%%%%%%%%%%%%%%%%%%%%%%%%%%%

\subsection{Flat directions during and after inflation}
\label{FDDAI}

Because it does not cost anything in energy during inflation, where
the Hubble expansion rate is $H_I \gg m_0$, quantum fluctuations are
free to accumulate (in a coherent state) along a flat direction and
form a condensate with a large VEV, $\varphi_0$.  Because inflation
smoothes out all gradients, only the homogeneous condensate mode
survives. However, the zero point fluctuations of the condensate
impart a small, and in inflationary models a calculable, spectrum of
perturbations on the condensate~\cite{flat}. In an abuse of language
we will collectively call such condensates as flat directions.

After inflation, $H \propto t^{-1}$, the flat direction stays at a
relatively larger VEV due to large Hubble friction term, note that the
Hubble expansion rate gradually decreases but it is still large
compared to $m_{0}$. When $H \simeq m_0$, the condensate along the
flat direction starts oscillating around the origin with an initial
amplitude $\sim \varphi_0$. From then on $\vert \varphi \vert$ is
redshifted by the Hubble expansion $\propto H$ for matter dominated
and $\sim H^{3/4}$ for radiation dominated Universe.

If higher order superpotential terms are forbidden, due to an
$R-$symmetry (or a set of $R-$symmetries)~\cite{dis}, then we
naturally have, $\varphi_0 \sim M_{\rm P}$~\cite{drt}.  On the other
hand, $\varphi_0 \ll M_{\rm P}$ will be possible if non-renormalizable
superpotential terms are allowed.  Let us first analyze the model
independent case where we treat $\varphi$ as a free parameter which
can vary in a wide range with an upper limit $\varphi_0 \leq M_{\rm
P}$, and study various consequences.  Note that a lower bound $H_I
\leq \varphi_0$ is set by the uncertainty due to quantum fluctuations
of $\varphi$ during inflation.

%%%%%%%%%%%%%%%%%%%%%%%%%%%%%%%%%%%%%%%%%%%%%%%%%%%%%%%%%%%%%%%%%%%%

\subsection{Flat directions and inflaton decay}
\label{FDID}

If a flat direction which has a non-zero VEV has couplings to the
inflaton decay product(s), then it will induce a mass, $y \vert
\varphi \vert$, where $y$ is a gauge or Yukawa coupling. The inflaton
decay at the leading order will be kinematically forbidden if $y \vert
\varphi \vert \geq m_{\phi}/2$. One should then wait until the Hubble
expansion has redshifted, $\vert \varphi \vert$, down to
$(m_{\phi}/2y)$. The decay happens when (note that $\vert \varphi
\vert \propto H$, after the flat direction starts oscillating and
before the inflaton decays):
\beq \label{dec1}
H_1 = {\rm min} ~ \left[\left({m_{\phi} \over y \varphi_0} \right) m_0 , 
\Gamma_{\rm d} \right].
\eeq
The inflaton also decays at higher orders of perturbation theory to
particles which are not directly coupled to it~\cite{ad1}. This mode
is kinematically allowed at all times, but the rate is suppressed by a
factor of $\sim \left(m_{\phi}/ y \vert \varphi \vert \right)^2
\Gamma_{\rm d}$. It will become efficient at
\beq \label{dec2}
H_2 \sim \left({m_{\phi} m_0 \over \varphi_0}\right)^{2/3} 
\Gamma^{1/3}_{\rm d}.
\eeq
Therefore, if the decay products are coupled to a flat direction
with a non-zero VEV, the inflaton will actually decay at a time when
the expansion rate of the Universe is given by
\beq \label{infdec}
H_{\rm d} = {\rm max} ~ \left[H_1, H_2 \right].
\eeq
In general it is possible to have, $H_{\rm d} \ll \Gamma_{\rm d}$,
particularly for large values of $\varphi_0$~\footnote{Flat directions
can also affect inflaton decay in other ways~\cite{abm}.}. Flat
directions can therefore significantly delay inflaton decay on purely
kinematical grounds~\footnote{Note that such a delayed decay can also
naturally implement the modulated fluctuations mechanism for
generating adiabatic density perturbations by converting their
isocurvature perturbations~\cite{modulated}.  Due to the dependence of
$H_{\rm d}$ on $\varphi_0$, fluctuations of the flat direction ($\sim
H_I$) imply an inhomogeneous inflaton decay which can give rise to
perturbations of the correct magnitude if, $H_I \sim 10^{-5}
\varphi_0$.}.
 
%%%%%%%%%%%%%%%%%%%%%%%%%%%%%%%%%%%%%%%%%%%%%%%%%%%%%%%%%%%%%%%%%%%%%%%%%%%%%%
\subsection{Flat directions and thermalization}\label{FDAT}

Flat directions can dramatically affect thermal history of the
Universe even if they do not delay the inflaton decay. The reason is
that the flat direction VEV spontaneously breaks the SM gauge
group. The gauge fields of the broken symmetries then acquire a
supersymmetry conserving mass, $m_{\rm g} \sim g \vert \varphi \vert$,
from their coupling to the flat direction, where $g$ is a gauge
coupling constant.

The simplest example is the flat direction corresponding to the $H_u
H_d$ monomial.  One can always rotate the field configuration to a
basis where $H_{u,1} = H_{d,2} = 0$, with subscripts $1$ and $2$
denoting the upper and lower components of the Higgs doublets
respectively.  The complex scalar field, $\varphi = \left(H_{u,2} +
H_{d,1}\right)/\sqrt{2}$, represents a flat direction. It breaks the
$SU(2)_W \times U(1)_Y$ down to $U(1)_{\rm em}$ (exactly in a similar
fashion as what happens in the electroweak vacuum). The $W^{\pm}$ and
$Z$ gauge bosons then obtain a mass from their couplings to the Higgs
fields via covariant derivatives.

The complex scalar field, $\left(H_{u,1} + H_{d,2} \right)/\sqrt{2}$,
and the real part of, $\left(H_{u,2} - H_{d,1}\right)/\sqrt{2}$, also
acquire the same mass as $W^{\pm}$ and $Z$, respectively, through the
$D-$term part of the scalar potential. The Higgsino fields
${\widetilde H}_{u,1}$ and ${\widetilde H}_{d,2}$ are paired with the
Winos, while $\left({\widetilde H}_{u,2} - {\widetilde
H}_{d,1}\right)/\sqrt{2}$ is paired with the Zino, and acquire the
same mass as $W^{\pm}$ and $Z$, respectively, through the
gaugino-gauge-Higgsino interaction terms.  In the supersymmetric
limit, the flat direction and its fermionic partner $\left({\widetilde
H}_{u,2} + {\widetilde H}_{d,1}\right)/\sqrt{2}$, as well as the
photon and photino, remain massless.

Many of the flat directions break the entire SM gauge group. The
prominent examples are flat directions corresponding to the $LLddd$
and $QuQue$ monomials. The whole SM gauge group can be also broken by
multiple flat directions which individually break only part of the
symmetry. For example flat directions corresponding to the $LLe$
monomial~\footnote{If the $LLe$ monomial develops a VEV during
inflation, which breaks the electroweak symmetry, it generates a mass
to the weak hypercharge gauge boson which mixes to the photon. It is
then possible to excite that gauge boson from vacuum fluctuations,
which would then be converted to the primordial magnetic field,
see~\cite{ejm}.} break the $SU(2)_W \times U(1)_Y$ group, while
directions corresponding to the $udd$ monomial break the
$SU(3)_C$. Note that all independent flat directions can
simultaneously acquire a large VEV during inflation~\footnote{Only
exceptional flat direction is the $H_u H_d$ direction. Since it has
superpotential Yukawa couplings to all MSSM fields, it will induce a
large mass to other scalar fields.  As a result no other flat
direction can develop a large VEV in the presence of $H_u H_d$. This
is not the case for the $H_u L$ flat direction: $LLddd$ has no Yukawa
couplings to it, and hence can simultaneously obtain a large VEV. For
a discussion on multiple flat directions, see~\cite{Asko-Maz}.}.
%The whole SM gauge group $SU(3)_C\times
%SU(2)_W \times U(1)_Y$ will be broken by the flat directions
%corresponding to the $LLddd$ and $QuQue$ monomials.

Let us now imagine one of such flat directions which has obtained a
large VEV during inflation. In which case a flat direction can
crucially alter thermal history of the Universe by suppressing
thermalization rate of the reheat plasma. Note that, $m_{\rm g}$,
provides a physical infrared cut-off for scattering diagrams with
gauge boson exchange in the $t-$channel shown in Figs.~(1,~2).  The
thermalization rate will then be given by (up to a logarithmic
``bremsstrahlung'' factor):
\beq \label{chemical1}
\Gamma_{\rm thr} \sim \alpha^2 {n \over {\vert \varphi \vert}^2},
\eeq
where we have used $m^2_{\rm g} \simeq \alpha {\vert \varphi
\vert}^2$. After the flat direction starts its oscillations at $H
\simeq m_0$, the Hubble expansion redshifts, ${\vert \varphi \vert}^2
\propto R^{-3}$, where $R$ is the scale factor of the FRW
Universe. The interesting point is that, $n \propto R^{-3}$, as well,
and hence $\Gamma_{\rm thr}$ remains constant while $H$ decreases for
$H < m_0$. This implies that $\Gamma_{\rm thr}$ eventually catches up
with the expansion rate, even if it is initially much smaller, and
shortly after that the full thermal equilibrium will be achieved.

Therefore the flat directions, even if they do not delay the inflaton
decay, will modify thermal history of the Universe. Depending on
whether $m_0 > \Gamma_{\rm d}$ or $m_0 < \Gamma_{\rm d}$, different
situations will arise which we discuss separately. If two or more flat
directions with non-zero VEVs induce mass to the gauge bosons, then
$\vert \varphi \vert$ denotes the largest VEV.

\begin{itemize}

\item{ $m_0 > \Gamma_{\rm d}$:\\ In this case the inflaton decays
after the flat direction oscillations start.  The inflaton
oscillations, which give rise to the equations of state close to
non-relativistic matter, dominate the energy density of the Universe
for $H > \Gamma_{\rm d}$. This implies that $R \propto H^{-2/3}$, and
$\vert \varphi \vert$ is redshifted $\propto H$ in this period. We
therefore have,
\beq \label{VEV1}
\varphi_{\rm d} \sim \left({\Gamma_{\rm d} \over m_0}\right) \varphi_0\,,
\eeq
where $\varphi_{\rm d}$ denotes the amplitude of the flat direction
oscillations at the time of the inflaton decay $H \simeq \Gamma_{\rm
d}$. The number density of particles in the reheat plasma at this time
is given by:
\beq \label{number1}
n_{\rm d} \sim {10 \Gamma^2_{\rm d} M^2_{\rm P} \over m_{\phi}}\,,
\eeq
where we have used $\rho \approx 3 \left(\Gamma_{\rm d} M_{\rm
P}\right)^2$.  Note that $\vert \varphi \vert$ and $n$ are both
redshifted $\propto R^{-3}$ for $H < \Gamma_{\rm d}$. Thus, after
using Eq.~(\ref{chemical1}), we find that complete thermalization
occurs when the Hubble expansion rate is
\beq \label{thermal1}
H_{\rm thr} \sim 10 \alpha^2 
%\left[{\left(m^{n-3}_0 M_{\rm P}\right)^{2/n-2} \over m_{\phi}}\right]
\left({M_{\rm P} \over \varphi_0}\right)^2 {m^2_0 \over m_{\phi}}\,.
\eeq
}

%%%%%%%%%%%%%%%%%%%%%%%%%%%%%%%%%%%%%%

\item{ $m_0 < \Gamma_{\rm d}$:\\ In this case the flat direction
starts oscillating after the completion of inflaton decay.  The
Universe is dominated by the relativistic inflaton decay products for
$H < \Gamma_{\rm d}$, implying that $R \propto H^{-1/2}$.

The number density of particles in the plasma is redshifted $\propto
H^{3/2}$ and, see Eq.~(\ref{number1}), at $H = m_0$, it is given by
\beq \label{number2}
n_0 \sim {10 \Gamma^2_{\rm d} M^2_{\rm P} \over m_{\phi}} 
\left({m_0 \over \Gamma_{\rm d}}\right)^{3/2}\,.
\eeq
Note that $n,~{\vert \varphi \vert}^2 \propto R^{-3}$, and hence
$\Gamma_{\rm thr}$ remains constant, for $H < m_0$. The reheat plasma
then fully thermalizes when the Hubble expansion rate is
\beq \label{thermal2}
H_{\rm thr} \sim 10 \alpha^2 
%\left[{\left(m^{n-3}_0 M_{\rm P}\right)^{2/n-2} \over m_{\phi}}\right] 
%\left({\Gamma_{\rm d} \over m_0}\right)^{1/2}
\left({\Gamma_{\rm d} \over m_0}\right)^{1/2} \left({M_{\rm P} 
\over \varphi_0}\right)^2 {m^2_0 \over m_{\phi}}\,.
\eeq
}

\end{itemize}

%%%%%%%%%%%%%%%%%%%%%%%%%%%%%%%%%%%%%%%%%%%%%%%

Note that the kinetic equilibrium is built through $2 \rightarrow 2$
scattering diagrams as in Fig.~(1), which have one interaction vertex
less than those in Fig.~(2). This implies that the rate for
establishment of kinetic equilibrium will be $\Gamma_{\rm kin} \sim
\alpha^{-1} \Gamma_{\rm thr}$.  Therefore we have a typical
relationship:
\begin{equation}
\Gamma_{\rm d} \gg \Gamma_{\rm kin} > \Gamma_{\rm thr}\,,
\end{equation}
in SUSY~\footnote{Relative chemical equilibrium among different
degrees of freedom is built through $2 \rightarrow 2$ annihilations in
the $s-$channel with a rate $\sim \alpha^2 n/E^2 \ll \Gamma_{\rm
thr}$. Hence composition of the reheat plasma will not change until
full thermal equilibrium is achieved.}.

This implies that the Universe enters a long period of quasi-adiabatic
evolution after the inflaton decay has completed. During this phase,
the comoving number density and (average) energy of particles remain
constant. It is only much later that the $2 \rightarrow 2$ and $2
\rightarrow 3$ scatterings become efficient and the plasma completely
thermalizes.  As we shall see, this has important cosmological
consequences.

One comment is in order before closing this section. So far we have
neglected the decay of the flat directions and their interactions with
the reheat plasma. These effects are considered in an
Appendix~\ref{additional}, and shown to be negligible before the
Universe fully thermalizes.

%%%%%%%%%%%%%%%%%%%%%%%%%%%%%%%%%%%%%%%%%%%%%%%%%%%%%%%%%%%%%%%%%%%%%%%%%%%%%%
%%%%%%%%%%%%%%%%%%%%%%%%%%%%%%%%%%%%%%%%%%%%%%%%%%%%%%%%%%%%%%%%%%%%%%%%%%%%%%
\section{Reheat temperature of the Universe}
\label{RTOTU}

The temperature of the Universe after it reaches full thermal
equilibrium is referred to as the reheat temperature $T_{\rm R}$. In
the case of SUSY, we therefore have:
\beq \label{rehtemp}
T_{\rm R} \simeq \left(H_{\rm thr} M_{\rm P}\right)^{1/2}\,,
\eeq
where, depending on the details, $H_{\rm thr}$ is given by
Eqs.~(\ref{thermal1}) and~(\ref{thermal2}).

As mentioned earlier, we typically have $H_{\rm thr} \ll \Gamma_{\rm
d}$. Therefore in SUSY theories the reheat temperature is generically
much smaller than the standard expression $T_{\rm R} \simeq
\left(\Gamma_{\rm d} M_{\rm P}\right)^{1/2}$, which is often used in
the literature and assumes immediate thermalization after the inflaton
decay.

An interesting point to note that the reheat temperature depends very
weakly on the inflaton decay rate, for instance Eq.~(\ref{thermal2})
implies that $T_{\rm R} \propto \Gamma^{1/4}_{\rm d}$, while $T_{\rm
R}$ is totally independent of $\Gamma_{\rm d}$ in
Eq.~(\ref{thermal1}). This is not difficult to understand, regardless
of how fast the inflaton decays, the Universe will not thermalize
until the $2 \rightarrow 3$ scatterings become efficient. The rate for
these scatterings essentially depends on the flat direction VEV and
mass.

As expected, a larger $\varphi_0$ results in slower thermalization and
a lower reheat temperature. On the other hand, larger values of $m_0$
lead to a higher $T_{\rm R}$. Since the flat direction oscillations
start earlier in this case, its VEV (thus mass of gauge bosons) is
redshifted faster and thermalization rate will be less suppressed.

The reheat temperature, $T_{\rm R}$, depends on the inflation sector
through the inflaton mass $m_{\phi}$. It is counterintuitive, but the
fact is that, from the above Eqs.~(\ref{thermal1}),~(\ref{thermal2})
and~(\ref{rehtemp}), a larger $m_{\phi}$ results in a lower reheat
temperature. This is due to the conservation of energy which implies
that the number density of inflaton decay products is inversely
proportional to their energy, which is $\sim m_{\phi}$
initially. Since the scattering rate is proportional to the number
density, a larger $m_{\phi}$ thus results in a smaller $\Gamma_{\rm
thr}$, and a lower $T_{\rm R}$.

If $H_{\rm thr}$ is very large or $\Gamma_{\rm d}$ is very small, we
may find $H_{\rm thr} \geq \Gamma_{\rm d}$. Obviously thermalization
cannot occur before the inflaton decay has completed. This merely
reflects the fact that the $2 \rightarrow 3$ scatterings are already
efficient when $H \simeq \Gamma_{\rm d}$.  This will be the case if
the flat direction VEV is sufficiently small at the time of the
inflaton decay, and/or if the reheat plasma is not very dilute.  The
former happens for a small $\varphi_0$ or large $m_0$, while smaller
values of $m_{\phi}$ lead to the latter case. The reheat temperature
in such cases follows the standard expression: $T_{\rm R} \simeq
\left(\Gamma_{\rm d} M_{\rm P}\right)^{1/2}$.

%%%%%%%%%%%%%%%%%%%%%%%%%%%%%%%%%%%%%%%%%%%%%%%%%%%%%%%%%%%%%%%%%%%%%%%%%%
\subsection{Numerical examples}\label{NE-1}

We now present some examples to demonstrate the impact of flat
directions on the rate of thermalization.  We choose the nominal value
of the inflaton mass, $m_{\phi} = 10^{13}$~GeV, typical value for the
flat direction mass in models with a weak scale SUSY, $m_0 \simeq
100~{\rm GeV}-1$ TeV, and four representative VEVs, $\varphi_0 =
M_{\rm P},~10^{-2} M_{\rm P},~10^{-4}M_{\rm P}$, and $\varphi_0 \leq
10^{-6} M_{\rm P}$.

If the inflaton decays gravitationally, see the example in
Appendix~\ref{hidden}, we have $\Gamma_{\rm d} \sim 10$ GeV. In this
case $H_{\rm thr}$ is given by Eq.~(\ref{thermal1}).  For larger
inflaton couplings to matter, $\Gamma_{\rm d} > m_0$ can be obtained,
in which case $H_{\rm thr}$ follows Eq.~(\ref{thermal2}). As a sample
case we have chosen $\Gamma_{\rm d} = 10^4$~GeV.

\vspace*{8mm}
\begin{center}
\begin{tabular}{|r|r|r|}
\hline
VEV (in GeV) & $T_{\rm R}(\Gamma_{\rm d} = 10$~GeV) & 
$T_{\rm R}(\Gamma_{\rm d} = 10^4$~GeV) \\
\hline \hline
$\varphi_0 \sim M_{\rm P}$ & $3 \times 10^3$ & $7 \times 10^{4}$ \\
\hline
$\varphi_0 = 10^{-2} M_{\rm P}$ & $3 \times 10^5$ & $7 \times 10^6$ \\
\hline
$\varphi_0 = 10^{-4} M_{\rm P}$ & $3 \times 10^7$ & $7 \times 10^8$ \\
\hline
$\varphi_0 \leq 10^{-6} M_{\rm P}$ & $3 \times 10^9$ & $7 \times 10^{10}$ \\
\hline
\end{tabular} 
\vspace*{2mm}
\end{center}
\noindent
{\bf Table 1:}~ The reheat temperature of the Universe for the
inflaton mass, $m_{\phi} = 10^{13}$ GeV, and two values of the
inflaton decay rate, $\Gamma_{\rm d} = 10,~10^4$ GeV.  The flat
direction mass is $m_0 \sim 1$~TeV. The rows show the values of
$T_{\rm R}$ for flat direction VEVs varying in a wide range.

\vspace*{6mm}
It is evident that thermalization rate (hence $T_{\rm R}$) becomes
smaller as $\varphi_0$ increases. It is remarkable that in the extreme
case where $\varphi_0 \sim M_{\rm P}$, the reheat temperature can be
as low as TeV. This is in stark contrast with the case if
thermalization was instant (as expected in a non-SUSY case), which
would result in hierarchically higher reheat temperatures $T_{\rm R}
> 10^{9}$ GeV (for the chosen values of $\Gamma_{\rm d}$).  Note that
thermalization still remains quite slow for $\varphi_0 \sim 10^{-4}
M_{\rm P}$.  It is only for $\varphi_0 \leq 10^{-6} M_{\rm P}$ that
the flat direction VEV is sufficiently small in order not to affect
thermalization, thus leading to the standard expression $T_{\rm R}
\simeq \left(\Gamma_{\rm d} M_{\rm P}\right)^{1/2}$.

This clearly underlines the fact that complete thermalization can be
substantially delayed in supersymmetry. Indeed, within the range
determined by the uncertainty of the quantum fluctuations, $m_{\phi}
\leq \varphi_0 \lsim M_{\rm P}$, flat directions considerably slow
down thermalization. This has important cosmological consequences
which will be considered in detail in the following sections.

%%%%%%%%%%%%%%%%%%%%%%%%%%%%%%%%%%%%%%%%%%%%%%%%%%%%%%%%%%%%%%%%%%
%%%%%%%%%%%%%%%%%%%%%%%%%%%%%%%%%%%%%%%%%%%%%%%%%%%%%%%%%%%%%%%%%%

\section{Thermalization in supersymmetric theories: Part-2}\label{part2}

So far we have not made any specific assumption on the mediation of
SUSY breaking to the observable sector, or the nature of higher order
terms which lift the flat directions. Instead we considered the
typical case in models with soft masses at TeV scale, and we treated
the flat direction VEV as a free parameter which is bounded from above
by $M_{\rm P}$. In this section we shall study issues and some
subtleties that may arise in more detail.

%%%%%%%%%%%%%%%%%%%%%%%%%%%%%%%%%%%%%%%%%%%%%%%%%%%%%%%%%%%%%%%%%%%%%%%%%%%%%
\subsection{Higher order superpotential terms and K\"ahler corrections}
\label{HOSTKC}

In models with gravity~\cite{nilles} and anomaly~\cite{anomaly}
mediation SUSY breaking results in as usual soft term, $m^2_0 {\vert
\varphi \vert}^2$, in the scalar potential where $m_0 \simeq 100~{\rm
GeV}-1$ TeV. There is also a new contribution arising from integrating
out heavy modes beyond the scale $M$, which usually induces
non-renormalizable superpotential term:
\begin{equation}
W\sim \lambda_{n}\frac{{\Phi}^n}{M^{n-3}}\,,
\end{equation}
where $\Phi$ denotes the superfield which comprises the flat direction
$\varphi$. In general $M$ could be the string scale, below which we
can trust the effective field theory, or $M = M_{\rm P}$ (in the case
of supergravity). In addition, there are also inflaton-induced
supergravity corrections to the flat direction potential. By
inspecting the scalar potential in $N=1$ supergravity, one finds the
following terms
\begin{eqnarray} \label{cross}
&&\left(e^{K(\varphi^{\ast},\varphi)/M_{\rm P}^2} V(I)\right)\,,~~~~
\left(K_{\varphi}K^{\varphi 
\varphi^{\ast}}K_{\varphi^{\ast}}\frac{{\vert W(I)\vert}^2}{M_{\rm P}^4}
\right)\,, ~~~~ \left(K_{\varphi}K^{\varphi I^{\ast}}
D_{I}\frac{W^{\ast}(I)}{M_{\rm P}^2}
+{\rm h.c.}\right)\, , \nonumber \\
&& \, 
\end{eqnarray}
where $D_{I}\equiv \partial/\partial I + K_{I}W/M_{\rm P}^2$. The
K\"ahler potential for the flat direction and the inflaton are given
by $K(\varphi^{\ast},\varphi)$ and $K(I^{\ast},I)$, and $W(I)$ denotes
the superpotential for the inflaton sector.

All these terms provide a general contribution to the flat direction
potential which is of the form~\cite{drt}:
\begin{equation}
\label{mflat}
V(\varphi)=H^2M_{\rm P}^2 f\left(\frac{\varphi}{M_{\rm P}}\right)\,,
\end{equation}
where $f$ is some function.  Such a contribution usually gives rise to
a Hubble induced correction to the mass of the flat direction with an
unknown coefficient, which depends on the nature of the K\"ahler
potential.  The relevant part of the scalar potential is then given
by~\cite{drt}
\beq \label{nonren}
V \supset \left(m^2_0 + c_H H^2 \right) {\vert \varphi \vert}^2 + 
\lambda_{n} {{\vert \varphi \vert}^{2(n-1)} \over M^{2(n-3)}}\,,
\eeq
with $n \geq 4$.  Note that $c_H$ can have either sign. If $c_H \gsim
1$, the flat direction mass is $> H$. It therefore settles at the
origin during inflation and remains there. Since $\vert \varphi \vert
= 0$ at all times, the flat direction will have no interesting
consequences in this case~\footnote{The positive Hubble induced mass
to the flat direction has a common origin to the Hubble induced mass
correction to the inflaton in supergravity models. This is the well
known $\eta$-problem~\cite{eta}, which arises because of the canonical
form of the inflaton part of the K\"ahler potential. A large $\eta$
generically spoils slow roll inflation. In order to have a successful
slow roll inflation, one needs $\eta \ll 1$ (and hence $c_{H}\ll 1$).
Note that the origin of the Hubble induced corrections to the mass of
flat direction is again the cross terms between inflaton K\"ahler
potential and superpotential terms and flat direction K\"ahler
potential. There is no absolutely satisfactory solution to the $\eta$
problem and therefore large coefficient, i.e., $c_{H}\sim +{\cal
O}(1)$. In this paper we assume that somehow the $\eta$ problem has
been addressed and therefore also $c_{H} \ll +{\cal O}(1)$.}. The case
with $c_H < 0$ will be more interesting. A negative $c_H$ can arise at
a tree-level~\cite{drt}, or as a result of radiative
correction~\cite{gmo,adm}. The flat direction VEV is in this case
driven away from the origin and quickly settles at a large value which
is determined by higher order terms that stabilize the potential.  A
large VEV can also be obtained if $0 < c_H \ll 1$.

It is possible to eliminate certain or all higher order superpotential
terms terms by assigning a suitable $R-$symmetry (or a set of
$R-$symmetries)~\cite{dis}.  However let us assume that all such terms
which respect the SM gauge symmetry are indeed present. As shown
in~\cite{gkm}, all of the MSSM flat directions are lifted by
higher-order terms with $n \leq 9$. If a flat direction is lifted at
the superpotential level $n$, the VEV that it acquires during
inflation will be: $\varphi_{\rm I} \sim \left(H_{I}
M^{n-3}\right)^{1/n-2}$, where $H_{I}$ is the expansion rate of the
Universe in the inflationary epoch.  If $c_H < 0$, this will be the
location of the minimum of the potential stabilized by the
higher-order term.

After inflation, the flat direction VEV slides down to an
instantaneous value: $\left(H(t) M^{n-3}\right)^{1/n-2}$. Once $H(t)
\simeq m_0$, soft SUSY breaking mass term in the potential takes over
and the flat direction starts oscillating around its origin with an
initial amplitude: $\varphi_0 \sim \left(m_0
M^{n-3}\right)^{1/n-2}$. If $M = M_{\rm P}$, we will have $\varphi_0
\sim 10^{10}$ GeV for $n = 4$, and $\varphi_0 \sim 10^{16}$ GeV for $n
= 9$. In particular, $\varphi_0 > 10^{14}$ GeV, for the flat
directions lifted at $n \geq 6$ levels. These directions affect
thermalization considerably and lower the reheat temperature of the
Universe as we discussed earlier, see Table.~1.

%%%%%%%%%%%%%%%%%%%%%%%%%%%%%%%%%%%%%%%%%%%%%%%%%%%%%%%%%%%%%%%%%%%%%%%%%%

\subsection{Models with gauge-mediated supersymmetry breaking}
\label{GMSB}

Our estimations of the thermalization time scale,
Eqs.~(\ref{thermal1}) and~(\ref{thermal2}), are valid if the flat
direction potential is quadratic $m^2_0 {\vert \varphi \vert}^2$ for
all field values (up to $M_{\rm P}$). However the situation is more
subtle in models with gauge-mediated SUSY breaking~\cite{gr}.

In these models there exists a sector with gauge interactions that
become strong at a scale $\Lambda_{\rm DSB} \ll M_{\rm P}$. This
induces a non-perturbative superpotential and leads to a non-zero
$F$-component, $\sim \Lambda^2_{\rm DSB}$, for a chiral superfield
which breaks supersymmetry. The gravitino mass will therefore be:
$m_{3/2} \sim {\Lambda}^2_{\rm DSB}/M_{\rm P}$. At a next step, SUSY
breaking is fed to a messenger sector with a mass scale $m_{\rm
mess}$.  Finally, the soft SUSY breaking parameters are induced in the
observable sector by integrating out the messenger fields, which have
some common gauge interactions with the observable sector, resulting
in soft scalar masses $m_0$.

As a result, the flat direction potential has the following forms:
\begin{eqnarray} \label{gmsbpot1}
V &=& m^2_0 {\vert \varphi \vert}^2 \hspace{5.5cm}\vert \varphi 
\vert \leq 
m_{\rm mess}  \,, \nonumber \\
V &\sim &\left(m_0 m_{\rm mess}\right)^2 ~ 
{\rm ln}\left({\vert \varphi 
\vert \over m_{\rm mess}}\right) ~~~~~ m_{\rm mess} 
\ll \vert 
\varphi \vert < {m_0 m_{\rm mess} \over m_{3/2}} \, , \nonumber \\
V &= & m^2_{3/2} {\vert \varphi \vert}^2 \hspace{5.2cm} 
\vert \varphi \vert \geq 
{m_0 m_{\rm mess} \over m_{3/2}}\, .
\end{eqnarray}
This behavior can be understood as follows. The flat direction
obtains its soft SUSY breaking mass by integrating out the messenger
fields. For $\vert \varphi \vert > M_{\rm mess}$ these fields acquire
a large mass $\propto \vert \varphi \vert$ from the flat direction
VEV, for details, see~\cite{dmm}. The flat direction mass, i.e.,
$\sqrt{V''}$, then decreases and the potential varies very slowly. At
very large field values, however, dominance of the contribution from
the gravity mediation recovers the conventional, ${\vert \varphi
\vert}^2$, dependence on the potential, although now the gravity
mediated soft SUSY breaking contribution is $m_{3/2}$ instead $m_0$.

Now let us find the thermalization time scale in models with gauge
mediation. For small values of $m_{3/2}$, typically arising in these
models, the flat direction starts its oscillations when $H = m_{3/2}$
and $V(\varphi_0) = m^2_{3/2} {\varphi}^2_0$.  Also, the oscillations
usually start after the inflaton has decayed, i.e., such that
$\Gamma_{\rm d} > m_{3/2}$.

Note that ${\vert \varphi \vert}^2$ and the number density, $n$, are
redshifted alike $\propto R^{-3}$ for $H < m_{3/2}$, and hence
$\Gamma_{\rm thr}$ remains constant, while $H$ decreases.  This
changes when $\vert \varphi \vert \lsim m_0 m_{\rm mess}/m_{3/2}$,
after that $\vert \varphi \vert$ is redshifted $\propto
R^{-3}$~\cite{dmm}, and hence $\Gamma_{\rm thr}$ increases $\propto
R^3$. The transition occurs when $H \sim {m_0 m_{\rm
mess}/\varphi_0}$.  Here we have assumed that the Universe is
radiation dominated, thus $R \propto H^{-1/2}$, for $H < \Gamma_{\rm
d}$. Due to a rapid increase of the ratio: $\Gamma_{\rm thr}/H$, the
reheat plasma reaches full thermal equilibrium when $\vert \varphi
\vert > m_{\rm mess}$. After using Eqs.~(\ref{chemical1})
and~(\ref{number2}), we find that the Hubble expansion rate at the
time of thermalization is then given by:
\beq \label{gmsbthermal2}
H_{\rm thr} \sim  \left(10 \alpha^2 \right)^{2/5} ~ \left({\Gamma_{\rm d} 
m^3_0 m^3_{\rm mess} \over m^2_{\phi}}\right)^{1/5} ~ \left({M^4_{\rm P} 
m^3_{3/2} \over {\varphi}^7_0}\right)^{1/5}\,.     
\eeq
The maximum impact on thermalization happens in models with low-energy
gauge mediation~\cite{dmm}, which can give rise to very light
gravitinos. In these models $\Lambda_{\rm DSB}$, $m_{\rm mess}$ and
$m_0$ are all related to each other by one-loop factors: $m_{\rm mess}
\sim {\Lambda}_{\rm DSB}/16 \pi^2$ and $m_0 \sim m_{\rm mess}/16
\pi^2$~\cite{dmm}. After using these relations and taking into account
that $m_{3/2} \simeq {\Lambda}^2_{\rm DSB}/M_{\rm P}$,
Eq.~(\ref{gmsbthermal2}) reads
\beq \label{lgmthermal}
H_{\rm thr} \sim \left(10^{-9} \alpha^2 \right)^{2/5} 
\left({\Gamma_{\rm d} m_{3/2} \over m^2_{\phi}}\right)^{1/5} 
\left({M_{\rm P} \over \varphi_0}\right)^{7/5} m_{3/2}\,.     
\eeq
The reheat temperature then follows: $T_{\rm R} \sim \left(H_{\rm thr}
M_{\rm P}\right)^{1/2}$.

For example, consider a model where $\Lambda_{\rm DSB} \sim 10^7$ GeV,
$m_{\rm mess} \sim 10^5$ GeV and $m_0 \sim 1$ TeV. The gravitino mass
in this model is $m_{3/2} \simeq 100$ KeV.  Note that the potential is
quadratic for $\vert \varphi \vert \leq 10^3$ GeV and $\vert \varphi
\vert \geq 10^{12}$ GeV, while being logarithmic in the intermediate
region.  Table.~2 summarizes the reheat temperature in this model for
similar values of $m_{\phi}$, $\Gamma_{\rm d}$ and $\varphi_0$ as in
Table.~1. What we find is that $T_{\rm R}$ is considerably lower in
this case.

\vspace*{8mm}
\begin{center}
\begin{tabular}{|r|r|r|}
\hline
VEV (in GeV) & $T_{\rm R}(\Gamma_{\rm d} = 10$~GeV) & $T_{\rm R}
(\Gamma_{\rm d} = 10^4$~GeV) \\
\hline \hline
$\varphi_0 \sim M_{\rm P}$ & $8 \times 10^2$ & $3 \times 10^3$ \\
\hline
$\varphi_0 = 10^{-2} M_{\rm P}$ & $ 2 \times 10^4$ & $6 \times 10^4$ \\
\hline
$\varphi_0 = 10^{-4} M_{\rm P}$ & $5 \times 10^5$ & $2 \times 10^5$ \\
\hline
$\varphi_0 = 10^{-6} M_{\rm P}$ & $10^7$  & $3 \times 10^7$ \\
\hline
\end{tabular} 
\vspace*{2mm}
\end{center}
\noindent
{\bf Table 2:}~The reheat temperature of the Universe in a model with
low-energy gauge mediation where the gravitino mass is $m_{3/2} = 100$
KeV.  The inflaton mass and its decay rate are the same as in
Table.~1. The rows show the values of $T_{\rm R}$ for flat direction
VEVs chosen the same as in Table.~1.
\vspace*{6mm}

To conclude, in models with gauge mediation the Universe thermalizes
even more slowly and the resulting reheat temperatures are even lower
compared to that of gravity mediation case. This is mainly due to the
fact that the flat directions start their oscillations much later in
this case.

%%%%%%%%%%%%%%%%%%%%%%%%%%%%%%%%%%%%%%%%%%%%%%%%%%%%%%%%%%%%%%%%%%%%%%%%%%%%%%
\subsection{Split supersymmetry}
\label{SS}

Now we consider the opposite situation where $m_0 \gg 1$ TeV.  This
happens in the recently proposed split SUSY
scenario~\cite{split}. This scenario does not attempt to address the
hierarchy problem. It allows the scalars (except the SM Higgs doublet)
to be very heavy, while keeping the gaugino and Higgsino fields
light. This removes problems with flavor changing and $CP-$violating
effects induced by scalars at one-loop level. On the other hand, it
preserves attractive features like supersymmetric gauge coupling
unification and a light LSP which can account for the dark matter.

Note that from Eqs.~(\ref{thermal1}) and~(\ref{thermal2}), the
thermalization time scale becomes shorter for larger values of $m_0$,
because when $m_0$ is larger the flat directions start oscillating
earlier. Their VEV and the induced masses for the gauge bosons will
therefore be redshifted faster in this case. Let us consider the
favored range of the soft scalar masses in split SUSY: $10^8~{\rm GeV}
\leq m_0 \leq 10^{13}$ GeV. This removes the flavor changing and
$CP-$violating effects and results in an acceptable gluino
lifetime~\cite{adgr}. For the inflaton mass $m_{\phi} = 10^{13}$ GeV,
we find $H_{\rm thr} > 100$ GeV for all values of $\varphi_0 \leq
M_{\rm P}$.

Therefore in the case of split SUSY, we always have 
\beq \label{rehsplit}
T_{\rm R} \geq 10^{10}~{\rm GeV}\,.
\eeq
This is fairly a robust prediction.

This might seem unacceptably high for thermal gravitino
production. However, models of split SUSY can also accommodate
gravitinos with a mass $m_{3/2} > 50$ TeV~\cite{adgr,okada}. Such
superheavy gravitinos decay before BBN~\cite{bbn}, and hence are not
subject to the tight bounds coming from BBN~\cite{cefo,kkm}. As a
matter of fact, the gravitino overproduction can in this case turn
into a virtue. The late decay of gravitinos, below the LSP freeze-out
temperature, can produce the correct dark matter abundance in a
non-thermal fashion~\cite{adgr,ajm,supermassive}.

Moreover, if produced abundantly, the gravitinos will dominate the
Universe. Their decay in this case dilutes the baryon asymmetry.
Consequently, larger parts of the parameter space will become
available for thermal leptogenesis and flat direction
baryogenesis~\cite{supermassive}.

We conclude that, due to larger scalar masses, thermalization is not
affected by the flat directions in split SUSY. However, for $m_{3/2} >
50$ TeV, having high reheat temperatures and thermal overproduction of
gravitinos can be a privilege rather than a
handicap~\cite{ajm,supermassive}.

%%%%%%%%%%%%%%%%%%%%%%%%%%%%%%%%%%%%%%%%%%%%%%%%%%%%%%%%%%%%%%%%%
%%%%%%%%%%%%%%%%%%%%%%%%%%%%%%%%%%%%%%%%%%%%%%%%%%%%%%%%%%%%%%%%%

\section{Particle production during thermalization}
\subsection{Quasi-adiabatic evolution of the Universe}
\label{deviation}

Right after the inflaton decay has completed the energy density of the
Universe is given by, $\rho \approx 3 \left(\Gamma_{\rm d} M_{\rm
P}\right)^2$, and the average energy of particles is $\langle E
\rangle \simeq m_{\phi}$~\footnote{For example, in a two-body decay of
the inflaton, we have exactly $E = m_{\phi}/2$.}. Deviation from full
equilibrium can be quantified by the parameter $''{\cal
A}''$~\cite{am}, where
\beq \label{a} {\cal A} \equiv {3 \rho \over T^4} \sim 10^4
\left({\Gamma_{\rm d} M_{\rm P} \over m^2_{\phi}}\right)^2.  \eeq
Here we define $T \approx \langle E \rangle/3$, in accordance with
full equilibrium. Note that in full equilibrium, see Eq.~(\ref{full}),
we have ${\cal A} \approx g_{\ast}$ ($=228.75$ in the MSSM).  On the other
hand, see Eq.~(\ref{finpert}), after the inflaton decay we have
$\Gamma_{\rm d} \ll m^2_{\phi}/M_{\rm P}$ which implies that ${\cal A}
\ll 228.75$. One can also associate parameter ${\cal A}_i \equiv 3
\rho_i/T^4$ to the $i-$th degree of freedom with the energy density
$\rho_i$ (all particles have the same energy $E$, and hence $T$, as
they are produced in one-particle decay of the inflaton). Note that
${\cal A} = \sum_{i}{{\cal A}_i}$, and in full equilibrium we have
${\cal A}_i \simeq 1$.

As we have discussed in Section~\ref{TST-1}, thermalization is very
slow in supersymmetry. As a result, the Universe enters a long phase
of quasi-adiabatic evolution during which the comoving number density
and comoving (average) energy of particles in the plasma remain
constant. Since particles are produced in one-particle inflaton decay,
the distribution is peaked around the average energy. The $2
\rightarrow 2$ scatterings (which become efficient shortly before
complete thermalization) smooth out the distribution. This implies
that the reheat plasma is in a quasi-thermal state which is not far
from kinetic equilibrium, but grossly deviated from chemical
equilibrium. In this period, which lasts until the $2 \rightarrow 3$
scatterings shown in Fig.~(2) become efficient, the Hubble expansion
redshifts $\rho_i \propto R^{-4}$, $n_i \propto R^{-3}$ and $T \propto
R^{-1}$. Therefore ${\cal A}$ and ${\cal A}_i$ remain constant
throughout this epoch.  Note that ${\cal A}$ depends on the total
decay rate of the inflaton $\Gamma_{\rm d}$ and its mass $m_{\phi}$
through Eq.~(\ref{a}).  While, ${\cal A}_i$ are determined by the
branching ratio for the inflaton decay to the $i-$th degree of
freedom~\footnote{We would like to thank Antonio Masiero for
highlighting this fact.}.  The composition of the reheat plasma is
therefore model-dependent before its complete thermalization. However,
some general statements can be made based on symmetry arguments:

\begin{itemize}
\item{For $CP$-conserving couplings, the inflaton decay produces the
same number of particles and anti-particles associated to a given
field.}

\item{For a singlet inflaton (which is the case in almost all models)
\footnote{Only exceptional model is an example of assisted
inflation~\cite{Liddle} driven by $N$ supersymmetric flat
directions~\cite{asko-assist} which are not gauge singlets.} the gauge
invariance implies that inflaton has equal couplings (thus branching
ratios) to the fields which are in an irreducible representation of
the gauge (sub)group. This holds in the presence of spontaneous
symmetry breaking via flat direction VEVs, as long as the particle
masses induced by coupling to the flat direction are (much) smaller
than that of the inflaton mass.  }

\item{If the inflaton mass is (much) larger than the soft SUSY
breaking masses in the observable sector, the inflaton decay produces
the same number of bosonic and fermionic components of matter
superfields~\footnote{This holds for a perturbative decay which is
relevant for the last stage of reheating. Non-perturbative inflaton
decay via preheating, which may happen at an initial stage, produces
bosons much more abundantly than fermions as a result of strong SUSY
breaking by large occupation numbers.}.  }

\end{itemize}

During the quasi-adiabatic evolution of the reheated plasma, i.e., for
$H_{\rm thr} < H < \Gamma_{\rm d}$, we have~\footnote{One can express
${\cal A}_i$ in terms of a {\it negative} chemical potential $\mu_i$,
where ${\cal A}_i = {\rm exp}\left(\mu_i/T \right)$. Note that for a
large negative chemical potential, i.e., in a dilute plasma, the Bose
and Fermi distributions are reduced to the Maxwell-Boltzmann distribution and 
give essentially the same result. The
assignment of a chemical potential merely reflects the fact that the
number of particles remains constant until the number-violating
reactions become efficient. It does not appear as a result of a
conserved quantity (such as baryon number) which is due to some
symmetry. Indeed, assuming that inflaton decay does not break such
symmetries, here we assign the same chemical potential to particles
and anti-particles.}
\beq \label{kin}
\rho_i = {\cal A}_i {3 \over \pi^2} T^4 ~ ~ , ~ ~ n_i = 
{\cal A}_i {1 \over \pi^2} T^4,
\eeq
and the Hubble expansion rate follows 
\beq \label{hubble}
H \simeq {\cal A}^{1/2} \left({T^2 \over 3 M_{\rm P}}\right).
\eeq
In this epoch $T$ varies in a range $T_{\rm min} \leq T \leq T_{\rm
max}$, where $T_{\rm max} \approx m_{\phi}/3$ is reached right after
the inflaton decay.  Because of complete thermalization, $T$ sharply
drops from $T_{\rm min}$ to $T_{\rm R}$ at $H_{\rm thr}$, where the
conservation of energy implies that
\beq \label{FR}
T_{\rm R} = \left({{\cal A} \over 228.75}\right)^{1/4} 
T_{\rm min},
\eeq
The (final) entropy density is given by $s = \left(2 \pi^2/45 \right)
g_{\ast} T^3_{\rm R}$, where we take $g_{\ast} = 228.75$.

%%%%%%%%%%%%%%%%%%%%%%%%%%%%%%%%%%%%%%%%%%%%%%%%%%%%%%%%%%%%%%%%%%%%%%%%%%%%%%

\subsection{Quasi-thermal particle production}
\label{QTPP}

An important cosmological phenomenon is the production of (un)stable
particles arising in theories beyond the SM in the early Universe. Let
us consider such a particle, $\chi$, with a mass, $m_{\chi}$, which is
weakly coupled to the (MS)SM fields. Through its couplings, $\chi$
will be inevitably produced in a plasma consisting of only the (MS)SM
particles. In full equilibrium, one can calculate the abundance of a
given particle by taking a thermal average of the relevant processes
(for example, see Ref.~\cite{kt}).  The only inputs required are the
particle mass and its couplings to the (MS)SM fields, plus the reheat
temperature $T_{\rm R}$. This leads to a lower bound on the relic
abundance which is {\it independent} from the details of the reheating
(except for $T_{\rm R}$).

Here we are interested in particle production during the quasi-thermal
phase of the Universe. Usually, scatterings are the most important
processes for production of $\chi$. The number density of $\chi$,
denoted by $n_{\chi}$, then obeys
\beq \label{chipro}
{\dot{n}}_{\chi} + 3 H n_{\chi} = \sum_{i \leq j} \langle v_{\rm rel} ~
\sigma_{ij \rightarrow \chi}\rangle n_i n_j. 
%+ \sum_{i} \Gamma_{i \rightarrow \chi}~ n_i.
\eeq
Here $n_i$ and $n_j$ are the number densities of the $i-$th and
$j-$th particles, $\sigma_{ij \rightarrow \chi}$ is the cross-section
for producing $\chi$ from scatterings of $i$ and $j$, and the sum is
taken over all fields which participate in $\chi$ production. Also
$\langle \rangle$ denotes averaging over the distribution. Note that
production will be Boltzmann suppressed if $T < m_{\chi}/3$.
Therefore, to obtain the total number of $\chi$ quanta produced from
scatterings, it will be sufficient to integrate the RH side
of~(\ref{chipro}) from the highest temperature down to $m_{\chi}/3$.
The physical number density of $\chi$ produced at a time $t_1$ will be
redshifted by a factor of $\left(R_2/R_1 \right)^3$ at any later time
$t_2$. If the Universe is filled by relativistic particles, $H =
\left(1/2t \right)$ and $R \propto H^{-1/2}$. The relic abundance of
$\chi$, normalized by the entropy density, $s$, will then be
\beq \label{chidens}
{n_{\chi} \over s} \sim 10^{-5} \left({228.75 \over 
{\cal A}}\right)^{5/4} 
%\left({228.75 \over g_{\ast}}\right)^{3/2} 
\sum_{i,j} 
\left[\int_{T_{\rm min}}^{T_{\rm max}}{{\cal A}_i {\cal A}_j 
{\langle v_{\rm rel}~\sigma_{ij \rightarrow 
{\chi}}\rangle}~ M_{\rm P}~ dT}\right]\,, 
%~ ~ ~ ({\rm quasi-thermal})
\eeq
where we have used Eq.~(\ref{FR}).

%%%%%%%%%%%%%%%%%%%%%%%%%%%%%%%%%%%%%%%%%%%%%%%%%%%%%%%%%%%%%%%%%%%%%%%%%%%%%%
\section{Gravitino production in a quasi-thermal case}
\label{GPQT}

As we discussed in Section 3, slow thermalization in SUSY generically
results in a low reheat temperature compatible with the BBN bounds on
thermal gravitino production. However gravitinos are also produced
during the quasi-thermal phase prior to a complete thermalization of
the reheat plasma.  There are various channels for the gravitino
production from scatterings of gauge, gaugino, fermion and sfermion
quanta. The scattering cross-section for all such processes is
$\propto 1/M^2_{\rm P}$, but the numerical coefficient depends on a
specific channel. In full thermal equilibrium, all degrees of freedom
(except for their bosonic or fermionic nature) have the same
occupation number, and hence obtaining the total production
cross-section is rather easy. However, as mentioned earlier, the
composition of the reheat plasma is model-dependent in a quasi-thermal
phase. Therefore calculating the total cross-section is more involved
in this case.

An important distinction is that we {\it only} need to consider
scatterings of fermions and/or sfermions for the following reasons;
the gauge and gaugino quanta have large masses $\sim \alpha^{1/2}
\varphi_{\rm d}$ (induced by the flat direction VEV) at a time most
relevant for the gravitino production, i.e., when $H \simeq
\Gamma_{\rm d}$, therefore, they decay to lighter fermions and
sfermions at a rate $\sim {\alpha}^{3/2} \varphi^2_{\rm
d}/m_{\phi}$. Here $\alpha^{3/2} \varphi_{\rm d}$ is the decay width
at the rest frame of gauge/gaugino quanta, and $\varphi_{\rm
d}/m_{\phi}$ is the time-dilation factor. The decay rate is $\gg
\Gamma_{\rm d}$, thus gauge and gaugino quanta decay almost instantly
upon production, and they will not participate in the gravitino
production~\footnote{The inverse decay happens at a rate $\sim
{\alpha} n/E^2$ and, as mentioned earlier, it is inefficient before
complete thermalization of the plasma. This is just a reflection of
the fact that the reheat plasma is out of chemical equilibrium for $H
> H_{\rm thr}$.}.  This is indeed welcome as scatterings which include
gauge and/or gaugino quanta in the initial state (particularly
scattering of two gluons) have the largest production
cross-section~\cite{moroi,thermal}.

{\it As a consequence, production of the helicity $\pm 1/2$ states
will not be enhanced in a quasi-thermal phase as scatterings with a
gauge-gaugino-gravitino vertex will be absent.}

Now let us discuss the relevant scattering processes and production
cross-sections in the quasi-thermal phase~\footnote{One might wonder
that scatterings of (s)fermions off the flat direction condensate
would be the dominant source as the latter carries a much larger
number of (zero-mode) quanta. However, these scatterings are
suppressed for the same reasons as discussed in
Appendix~\ref{additional}.}.  The total cross-section, and
cross-sections for multiplets comprising the LH (s)quarks $Q$, RH
up-type (s)quarks $u$, RH down-type (s)quarks $d$, LH (s)leptons $L$,
RH (s)leptons $e$ and the two Higgs/Higgsino doublets $H_u,H_d$ have
been given separately for each channel~\footnote{We also note that
these cross-sections are free from infrared logarithmic
divergences~\cite{moroi,thermal}.}. Here $1 \leq i,j \leq 3$, $a,b =
1,2$ and $1 \leq \alpha, \beta \leq 3$ are the flavor, weak-isospin
and color indices of scattering degrees of freedom respectively. Also
$\alpha_3$, $\alpha_2$ and $\alpha_1$ are the gauge fine structure
constants associated to the $SU(3)_C$, $SU(2)_W$ and $U(1)_Y$ groups
respectively.  For each channel we have taken the total cross-section
from the table on page [181] of the first reference in~\cite{thermal},
and split it according to the gauge interactions for the relevant
flavors (and each of the two doublets in the case of Higgs). Note that
the prefactor $3/4$, which arises as a result of Fermi-Dirac
statistics in full thermal equilibrium will not be relevant here.
\begin{itemize}
\item{{\it fermion} + {\it anti-sfermion $\rightarrow$ gravitino} + {\it gauge 
field},\\
\noindent 
{\it sfermion} + {\it anti-fermion $\rightarrow$ gravitino} + 
{\it gauge field}:
The total cross-section for this channel is $\left(1/32 M^2_{\rm
P}\right) \times \left(48 \alpha_3 + 21 \alpha_2 + 11
\alpha_1\right)$, see row (E) of the table on page [181] of the first
reference in~\cite{thermal}. For different degrees of freedom we then
find:
\begin{eqnarray}  \label{sferfer} 
\sigma_1 = \sigma_2 = {1 \over 32 M^2_{\rm P}} \delta_{ij} \times & \, 
\nonumber \\
\left({2 \over 9} \alpha_3 \delta_{ab}+ {3 \over 16} \alpha_2 
\delta_{\alpha \beta} + {1 \over 72} \alpha_1 \delta_{\alpha \beta} 
\delta_{ab}\right) ~ ~ ~ ~ ~ & ~ ~ ~ ~ ~ 
Q \, , \nonumber \\
\left({2 \over 9} \alpha_3 \delta_{ab}+ {2 \over 9} \alpha_1 
\delta_{\alpha \beta} \right) ~ ~ ~ ~ ~ & ~ ~ ~ ~ ~ 
u \, , \nonumber \\
\left({2 \over 9} \alpha_3 \delta_{ab}+ {1 \over 18} \alpha_1 
\delta_{\alpha \beta} \right) ~ ~ ~ ~ ~ & ~ ~ ~ ~ ~ 
d \, , \nonumber \\
\left({3 \over 16} \alpha_2 + {1 \over 8} \alpha_1 
\delta_{ab}\right) ~ ~ ~ ~ ~ & ~ ~ ~ ~ ~ 
H \, , \nonumber \\
\left({3 \over 16} \alpha_2 + {1 \over 8} \alpha_1 
\delta_{ab}\right) ~ ~ ~ ~ ~ & ~ ~ ~ ~ ~ 
L \, , \nonumber \\
\left({1 \over 2} \alpha_1 \right) ~ ~ ~ ~ ~ & ~ ~ ~ ~ ~ 
e \, .
\end{eqnarray}
}    
\item{{\it fermion} + {\it anti-fermion} $\rightarrow$ {\it gravitino}
+ {\it gaugino}: The total cross-section for this channel is
$\left(1/32 M^2_{\rm P}\right) \times \left(16 \alpha_3 + 7 \alpha_2 +
11 \alpha_1/3\right)$, see row (I) of the table on page [181] of the
first reference in~\cite{thermal}. For different degrees of freedom we
then find:
\begin{eqnarray}  \label{ferfer} 
\sigma_3 = {1 \over 32 M_{\rm P}} \delta_{ij} \times & \, \nonumber \\
\left({4 \over 27} \alpha_3 \delta_{ab}+ {1 \over 8} \alpha_2 
\delta_{\alpha \beta} + {1 \over 108} \alpha_1 \delta_{\alpha \beta} 
\delta_{ab}\right) ~ ~ ~ ~ ~ & ~ ~ ~ ~ ~ 
Q \, , \nonumber \\
\left({4 \over 27} \alpha_3 \delta_{ab}+ {4 \over 27} \alpha_1 
\delta_{\alpha \beta} \right) ~ ~ ~ ~ ~ & ~ ~ ~ ~ ~ 
u \, , \nonumber \\
\left({4 \over 27} \alpha_3 \delta_{ab}+ {1 \over 27} \alpha_1 
\delta_{\alpha \beta} \right) ~ ~ ~ ~ ~ & ~ ~ ~ ~ ~ 
d \, , \nonumber \\
\left({1 \over 8} \alpha_2 + {1 \over 12} \alpha_1 
\delta_{ab}\right) ~ ~ ~ ~ ~ & ~ ~ ~ ~ ~ 
H \, , \nonumber \\
\left({1 \over 8} \alpha_2 + {1 \over 12} \alpha_1 
\delta_{ab}\right) ~ ~ ~ ~ ~ & ~ ~ ~ ~ ~ 
L \, , \nonumber \\
\left({1 \over 3} \alpha_1 \right) ~ ~ ~ ~ ~ & ~ ~ ~ ~ ~ 
e \, .
\end{eqnarray}
} 
\item{{\it sfermion} + {\it anti-sfermion} $\rightarrow$ {\it
gravitino} + {\it gaugino}: The total cross-section for this channel
is $\left(1/32 M^2_{\rm P}\right) \times \left(8 \alpha_3 + 7
\alpha_2/2 + 11 \alpha_1/6\right)$, see row (J) of the table on page
[181] of the first reference in~\cite{thermal}. For different degrees
of freedom we then find:
\begin{eqnarray}  \label{ferfer} 
\sigma_4 = {1 \over 32 M^2_{\rm P}} \delta_{ij} \times & \, \nonumber \\
\left({2 \over 27} \alpha_3 \delta_{ab}+ {1 \over 16} \alpha_2 
\delta_{\alpha \beta} + {1 \over 216} \alpha_1 \delta_{\alpha \beta} 
\delta_{ab}\right) ~ ~ ~ ~ ~ & ~ ~ ~ ~ ~ 
Q \, , \nonumber \\
\left({2 \over 27} \alpha_3 \delta_{ab}+ {2 \over 27} \alpha_1 
\delta_{\alpha \beta} \right) ~ ~ ~ ~ ~ & ~ ~ ~ ~ ~ 
u \, , \nonumber \\
\left({2 \over 27} \alpha_3 \delta_{ab}+ {1 \over 54} \alpha_1 
\delta_{\alpha \beta} \right) ~ ~ ~ ~ ~ & ~ ~ ~ ~ ~ 
d \, , \nonumber \\
\left({1 \over 16} \alpha_2 + {1 \over 24} \alpha_1 
\delta_{ab}\right) ~ ~ ~ ~ ~ & ~ ~ ~ ~ ~ 
H \, , \nonumber \\
\left({1 \over 16} \alpha_2 + {1 \over 24} \alpha_1 
\delta_{ab}\right) ~ ~ ~ ~ ~ & ~ ~ ~ ~ ~ 
L \, , \nonumber \\
\left({1 \over 6} \alpha_1 \right) ~ ~ ~ ~ ~ & ~ ~ ~ ~ ~ 
e \, .
\end{eqnarray}
}
\end{itemize}   
The total cross-section for the gravitino production will then be
given by:
\begin{eqnarray}  \label{ferfer} 
\sigma_{\rm tot} = {1 \over 32 M^2_{\rm P}} \delta_{ij} \times & \, 
\nonumber \\
\left({2 \over 3} \alpha_3 \delta_{ab}+ {9 \over 16} \alpha_2 
\delta_{\alpha \beta} + {1 \over 24} \alpha_1 \delta_{\alpha \beta} 
\delta_{ab}\right) ~ ~ ~ ~ ~ & ~ ~ ~ ~ ~ 
Q \, , \nonumber \\
\left({2 \over 3} \alpha_3 \delta_{ab}+ {2 \over 3} \alpha_1 
\delta_{\alpha \beta} \right) ~ ~ ~ ~ ~ & ~ ~ ~ ~ ~ 
u \, , \nonumber \\
\left({2 \over 3} \alpha_3 \delta_{ab}+ {1 \over 6} \alpha_1 
\delta_{\alpha \beta} \right) ~ ~ ~ ~ ~ & ~ ~ ~ ~ ~ 
d \, , \nonumber \\
\left({9 \over 16} \alpha_2 + {3 \over 8} \alpha_1 
\delta_{ab}\right) ~ ~ ~ ~ ~ & ~ ~ ~ ~ ~ 
H \, , \nonumber \\
\left({9 \over 16} \alpha_2 + {3 \over 8} \alpha_1 
\delta_{ab}\right) ~ ~ ~ ~ ~ & ~ ~ ~ ~ ~ 
L \, , \nonumber \\
\left({3 \over 2} \alpha_1 \right) ~ ~ ~ ~ ~ & ~ ~ ~ ~ ~ 
e \, .
\end{eqnarray}
As we pointed out in the previous section, particles and
anti-particles associated to the bosonic and fermionic components of
the multiplets which belong to an irreducible representation of a
gauge group have the same parameter ${\cal A}_i$. This implies that
\begin{eqnarray} \label{final} 
\Sigma_{\rm tot} \equiv \sum^{3}_{i,j=1} ~ \sum^{2}_{a,b=1} ~ 
\sum^{3}_{\alpha, \beta = 1} 
{{\cal A}_{i,a,\alpha} {\cal A}_{j, b, \beta} ~  
\langle \sigma_{\rm tot} v_{\rm rel} \rangle} = \, \nonumber \\
{1 \over 32 M^2_{\rm P}} \sum {[6 \alpha_3 (2 {\cal A}^2_{Q} + 
{\cal A}^2_{u} + {\cal A}^2_{d}) + 
{9 \over 4} \alpha_2 (3 {\cal A}^2_{Q} + {\cal A}^2_{L} + {\cal A}^2_{H})} 
\, \nonumber \\
{+ {1 \over 4} \alpha_1 ({\cal A}^2_{Q} + 8 {\cal A}^2_{u} + 2 
{\cal A}^2_{d} + 3 {\cal A}^2_{L} + 6 {\cal A}^2_{e} + 3 {\cal A}^2_{H})}] \, .
\end{eqnarray}
The sum is taken over the three flavors of $Q,u,d,L,e$ and the two
Higgs doublets.  After replacing $\Sigma_{\rm tot}$ in
Eq.~(\ref{chidens}), and recalling that $T_{\rm max} \approx
m_{\phi}/3$, we obtain
\beq \label{gravquasi}
{n_{3/2} \over s} \simeq \left(10^{-1} M^2_{\rm P} ~ \Sigma_{\rm tot}\right) 
\left({228.75 \over {\cal A}}\right)^{5/4} 
\left({T_{\rm max} \over 10^{10}~{\rm GeV}}\right) ~ 10^{-12}\,.
\eeq
We remind that in full thermal equilibrium, $\Sigma_{\rm tot} =
\left(4 \pi/M^2_{\rm P}\right) \times \left(16 \alpha_3 + 6 \alpha_2 +
2 \alpha_1 \right) \simeq \left(10^{-1}/M^2_{\rm P}\right)$ (up to
logarithmic corrections)~\cite{moroi,thermal}.
%Also note that the logarithmic corrections are due to renormalization group 
%evolution of gauge couplings
%
%\begin{eqnarray} \label{rg}
%\alpha_3 (\Lambda) \simeq 0.06 \left(1 - 0.02 {\rm ln} \left({\Lambda \over 
%10^9~{\rm GeV}}\right)\right) \, \nonumber \\
%\alpha_2 (\Lambda) \simeq 0.06 \left(1 + 0.007 {\rm ln} \left({\Lambda \over 
%10^9~{\rm GeV}}\right)\right) \, \nonumber \\    
%\alpha_1 (\Lambda) \simeq 0.06 \left(1 + 0.02 {\rm ln} \left({\Lambda \over 
%10^9~{\rm GeV}}\right)\right) \, \nonumber \\
%\end{eqnarray}
%
It is evident that the exact abundance of the gravitinos produced
during the quasi-thermal phase depends on the composition of the
reheat plasma. One expects the number of gravitinos thus produced to
be maximum if the inflaton mainly decays to one flavor of LH (s)quarks
(which are charged under the whole SM group).  In this case ${\cal
A}_Q = 1/24$ for the relevant flavor~\footnote{The total number of
degrees of freedom in one flavor of LH (s)quarks is $2~({\rm
particle-antiparticle}) \times 2~({\rm fermion-boson}) \times 2~({\rm
weak-isospin}) \times 3~({\rm colour})$.}, while ${\cal A} = 0$ for
all other degrees of freedom. This results in a gravitino abundance:
\begin{eqnarray} \label{gravmax} 
{n_{3/2} \over s} \simeq  \left({{\cal A} \over 228.75}\right)^{3/4} ~
\left({T_{\rm max} \over 10^{10}~{\rm GeV}}\right) ~ 10^{-12}\, , 
\end{eqnarray}
where ${\cal A}$ is given by Eq.~(\ref{a}).

An important point to note is that $T_{\rm max}$ is accompanied by the
factor ${\cal A}^{3/4}$ in Eq.~(\ref{gravmax}). Therefore, despite the
fact that $T_{\rm max} \approx m_{\phi}/3$ can be as large as
$10^{12}$ GeV, the gravitino abundance can be at a safe level.  First
consider the case for unstable gravitinos.  For $T_{\rm max} \simeq
10^{12}$ GeV, the tightest bound from BBN $\left(n_{3/2}/s\right) \leq
10^{-16}$ (arising for $m_{3/2} \simeq 1$ TeV and a hadronic branching
ratio $\simeq 1$) is satisfied if ${\cal A} \leq 10^{-6}$. Much weaker
bounds on ${\cal A}$ are found for a radiative decay. For example,
${\cal A} \leq 10^{-3}$ ($1$) if $m_{3/2} \simeq 100$ GeV ($1$ TeV).

For stable gravitinos with ${\cal O}({\rm keV})$ mass the dark matter
limit, see Eq.~(\ref{dmlimit}), is satisfied if ${\cal A} \leq
10^{-6}$.  A much more relaxed bound ${\cal A} < 1$ is obtained if
$m_{3/2} \simeq 100$ MeV. Therefore, in general, the gravitino
production during the quasi-thermal phase is safe~\footnote{The above
mentioned constraints on ${\cal A}$ are comfortably satisfied in a
generic inflationary model where the inflaton is a gauge singlet and
couples to the MS(SM) gravitationally, see the discussion in an
Appendix~\ref{hidden}, and in~\ref{sneutrino}.}.

{\it We conclude that late thermalization of the Universe due to SUSY
flat directions eliminates the gravitino problem altogether in a
natural way.}

%%%%%%%%%%%%%%%%%%%%%%%%%%%%%%%%%%%%%%%%%%%%%%%%%%%%%%%%%%%
%%%%%%%%%%%%%%%%%%%%%%%%%%%%%%%%%%%%%%%%%%%%%%%%%%%%%%%%%%%
\section{Leptogenesis in a quasi-thermal case}
\label{LQT}

\subsection{Basic concept}

The baryon asymmetry of the Universe (BAU) parameterized as $\eta_{\rm
B}=(n_{\rm B}-n_{\bar{\rm B}})/s$ is determined to be $0.9 \times
10^{-10}$ by the recent analysis of WMAP data~\cite{wmap}. This number
is also in good agreement with an independent determination from the
primordial abundance of light elements produced during
BBN~\cite{cfo}. Any mechanism for generating a baryon asymmetry must
satisfy: $B-$ and/or $L-$violation, $C-$ and $CP-$violation, and
departure from thermal equilibrium~\cite{sakharov}~\footnote{Since
$B+L$-violating sphaleron transitions are active at temperatures
$100~{\rm GeV} \lsim T \lsim 10^{12}$ GeV~\cite{krs}, any mechanism
for creating a baryon asymmetry at $T > 100$ GeV must create a $B-L$
asymmetry. The final asymmetry is then given by $B=a(B-L)$, where
$a=28/79$ in the case of SM and $a=8/23$ for the
MSSM~\cite{khlebnikov}.}.

Leptogenesis is an elegant mechanism which postulates the existence of
RH neutrinos, which are SM singlets, with a lepton number violating
Majorana mass $M_N$. It can be naturally embedded in models which
explain the light neutrino masses via the see-saw
mechanism~\cite{seesaw}. A lepton asymmetry can then be generated from
the out-of-equilibrium decay of the RH neutrinos into Higgs bosons and
light leptons, provided $CP-$violating phases exist in the neutrino
Yukawa couplings~\cite{fy,luty,one-loop}.  The created lepton
asymmetry will be converted into a baryonic asymmetry via sphaleron
processes.

In thermal leptogenesis the on-shell RH neutrinos whose decay is
responsible for the lepton asymmetry are produced via their Yukawa
interactions with the SM fields in a thermal bath~\cite{plumacher}. In
SUSY there is RH sneutrinos which serve an additional source for
leptogenesis~\cite{cdo}.  This scenario works most comfortably if
$T_{\rm R} \gsim M_{1} \geq 10^{9}$~GeV~\cite{buchmuller,sacha,gnrrs}
\footnote{There exist various scenarios of non-thermal
leptogenesis~\cite{reh,gprt,off-shell,bdps,my,bmp,adm1,ad3} which can
work for $T_{\rm R} \leq M_N$. There are also leptogenesis models
which implement soft SUSY breaking terms~\cite{soft}.}.

The decay of a RH (s)neutrino with mass $M_i$ results in a lepton
asymmetry via one-loop self-energy and vertex corrections
~\cite{one-loop}.  If the asymmetry is mainly produced from the decay
of the lightest RH states, and assuming hierarchical RH (s)neutrinos
$M_1 \ll M_2,M_3$, we will have~\cite{di}
\begin{eqnarray} \label{baryontherm}
\eta_{\rm B} & \simeq & 3 \times 10^{-10} 
\kappa  \left({m_3 - m_1 \over 0.05~{\rm eV}}\right) 
\left({M_1 \over 10^9~{\rm GeV}}\right)\,, \nonumber \\
 & & \hspace{8cm} ({\rm full ~ equilibrium})\,,
\end{eqnarray}
for ${\cal O}(1)$ $CP$-violating phases ($m_1 < m_2 < m_3$ are the
masses of light mostly LH neutrinos). Here $\kappa$ is the efficiency
factor accounting for the decay, inverse decay and scattering
processes involving the RH states~\cite{buchmuller,gnrrs}.

A decay parameter $K$ can be defined as
\beq \label{decpar}
K \equiv {\Gamma_1 \over H(T = M_1)}\,,
\eeq
where $\Gamma_1$ is the decay width of the lightest RH (s)neutrino. It
can be related to an effective neutrino mass ${\widetilde m}_1$ such
that $K = {\tilde m}_1/(10^{-3} ~ {\rm eV})$, with the
model-independent bound $m_1 < {\tilde m}_1$~\cite{fhy}.

If $K < 1$, corresponding to ${\tilde m}_1 < 10^{-3}$ eV, the decay of
the RH states will be out of equilibrium at all times. In this case
the RH states, which are mainly produced via scatterings of the LH
(s)leptons off the top (s)quarks and electroweak gauge/gaugino
fields~\cite{plumacher}, never reach thermal equilibrium.  The
cross-section for producing the RH (s)neutrinos is $\propto
T^{-2}~(M^2_1)$, when $T > M_1~(< M_1)$, and hence most of them are
produced when $T \sim M_1$.  The efficiency factor reaches its maximum
value for $\kappa \simeq 0.1$ when ${\tilde m}_1 = 10^{-3}$~eV. For
larger values of ${\widetilde m}_1$ it drops again, because the
inverse decays become important and suppress the generated
asymmetry. Producing sufficient asymmetry then sets a lower bound,
$M_1 \geq 10^9$ GeV~\cite{buchmuller}.  Successful thermal
leptogenesis therefore requires that $T_{\rm R} \geq 10^9$~GeV. Note
that this is at best marginally compatible with thermal gravitino
production, see Eq.~(\ref{gravtherm}).

%%%%%%%%%%%%%%%%%%%%%%%%%%%%%%%%%%%%%%%%%%%%%%%%%%%%%%%%%%%%%%%%%
\subsection{Quasi-thermal leptogenesis}
\label{QTL}

The presence of flat directions slow down thermalization and lower the
reheat temperature, such that, $T_{\rm R} \ll 10^9$ GeV is naturally
obtained (see Tables.~1 and~2). This is detrimental to thermal
leptogenesis in the bulk of parameter space. Note that
Eq.~(\ref{baryontherm}) implies that sufficient asymmetry will not be
generated after the establishment of a full
equilibrium~\footnote{Thermal leptogenesis can work for $M_1 \ll
10^9$~GeV if the RH (s)neutrinos are degenerate~\cite{pilaftsis}, or
for specific neutrino mass models~\cite{strumia}.}.  Needless to
mention, like gravitino production, leptogenesis can still occur
during the quasi-thermal phase and this is the topic of our interest
in this subsection.

In a quasi-thermal phase the reheat plasma is dilute, implying that
the RH states are produced less abundantly than in full thermal
equilibrium.  Their abundance can be calculated from
Eq.~(\ref{chidens}), where the production cross-section is, $\sigma
\propto 1/M^2_1$.  Note that the abundance depends on ${\cal A}_i$,
and hence on the composition of the reheat plasma. Significant
production of the RH (s)neutrinos requires that the LH (s)leptons
and/or the top (s)quarks be present in the reheat plasma. The relevant
channels for producing the lightest RH neutrino, $N_1$, and sneutrino,
${\tilde N}_1$, are scatterings of LH (s)leptons with the largest
Yukawa coupling to $N_1,{\tilde N}_1$ off the LH top (s)quarks and
anti-(s)quarks, and annihilation of top (s)quark-anti(s)quark
pairs~\footnote{The electroweak gauge/gaugino fields have a large mass
and decay almost instantly, therefore, unlike the case with full
equilibrium they do not participate in $N_1,{\tilde N}_1$
production.}. Summing over all processes (including weak-isospin and
color indices) we find from Eq.~(\ref{chidens}) that:
\beq \label{nprod}
{n_{N_1} \over s} \simeq 10^{-5} \Sigma_{N_1} \left(M_{\rm P} M_1\right) 
\left({228.75 \over {\cal A}}\right)^{5/4}\,, 
\eeq
where
\beq \label{sigman}
\Sigma_{N_1} \propto {1 \over M^2_1} \times 
18 \left({\cal A}_L {\cal A}_{Q_3} + {\cal A}_L {\cal A}_{t} + 
{\cal A}_{Q_3} {\cal A}_{t} \right)\,,
\eeq
is the production cross-section for $N_1,{\tilde N}_1$. Here ${\cal
A}_L$, ${\cal A}_{Q_3}$ and ${\cal A}_t$ denote the ${\cal A}$
parameter for the LH (s)leptons with the largest Yukawa coupling to
$N_1,{\tilde N}_1$, the LH top (s)quarks and the RH top (s)quarks
respectively. Note that in full equilibrium they are all $\simeq
1$. The three terms inside the parentheses are the contributions from
the above mentioned processes respectively. The final baryon asymmetry
generated in the quasi-thermal phase will then be given by:
\begin{eqnarray} \label{baryonquasi}
\eta_{\rm B} & \simeq & 10^{-10} \left({228.75 \over {\cal A}}\right)^{5/4} 
%\left({228.75 \over g_{\ast}}\right)^{3/2}  
\left({\cal A}_L {\cal A}_{Q_3} + {\cal A}_L {\cal A}_t + {\cal A}_{Q_3} 
{\cal A}_t \right)  \kappa  
\left({m_3 - m_1 \over 0.05~{\rm eV}}\right)  
\left({M_1 \over 10^9~{\rm GeV}}\right) \,. \nonumber \\ 
 & & \, \nonumber \\
 & & \hspace{8cm} ({\rm quasi-thermal})
\end{eqnarray}
If the inflaton mainly decays to the top (s)quarks, we have ${\cal
A}_{Q_3} = {\cal A}_t = {\cal A}/36$~\footnote{Note that the total
number of degrees of freedom in $Q_3$ and $t$ is 36.}, while ${\cal A}
= 0$ for all other degrees of freedom. This results in:
\beq \label{baryonmax}
\eta_{\rm B} \simeq 5 \times 10^{-9} \kappa 
\left({{\cal A} \over 228.75}\right)^{3/4} 
\left({M_1 \over 10^9~{\rm GeV}}\right)\,.
\eeq
Since $\eta_{\rm B} \propto M_1$, the maximum asymmetry is produced
when $M_1 \simeq 3 T_{\rm max} \approx 3 \times 10^{12}$ GeV. For the
largest efficiency factor $\kappa \simeq 0.1$, generating the correct
asymmetry requires that ${\cal A} \geq 10^{-3}$. This is compatible
with the bound from the gravitino production for a radiatively
decaying gravitino with $m_{3/2}\simeq 100~{\rm GeV}-1$ TeV (see the
discussion in the previous section, Eq.~(\ref{gravmax}).).

Note that from Eqs.~(\ref{gravmax}) and~(\ref{baryonquasi}) the
gravitino abundance and the baryon asymmetry produced during a
quasi-thermal phase are both $\propto {\cal A}^{3/4}$. However, due to
the dependence on the composition of the reheat plasma, the
marginality between leptogenesis and gravitino production can be very
different from that in the case of full equilibrium. The best case
scenario for leptogenesis occurs when the inflaton mainly decays to
the top (s)quarks. In this case the marginality between leptogenesis,
Eq.~(\ref{baryonmax}), and gravitino production, Eq.~(\ref{gravmax}),
is weakened by about one order of magnitude compared to the case of
full equilibrium, see Eqs.~(\ref{gravtherm}) and~(\ref{baryontherm}).

There are other interesting differences which arise in the case of a
quasi-thermal leptogenesis. Because the plasma is dilute in this
phase, $N_1,~{\tilde N}_1$ will not be brought into equilibrium even
if ${\widetilde m}_1 \gg 10^{-3}$ eV. On the other hand, see
Eq.~(\ref{hubble}), the expansion rate of the Universe is (much)
slower than the case with full thermal equilibrium when $T \simeq
M_1$. This implies that out-of-equilibrium decay of $N_1,~{\tilde
N}_1$ requires that ${\widetilde m}_1 \ll 10^{-3}$ eV.  Moreover,
since $\Delta L = 2$ scatterings mediated by $N_1,~{\tilde N}_1$ are
much less efficient in a dilute plasma, the resulting bound on $M_1$
will be altered.

{\it To conclude, late thermalization of the Universe implies that thermal
leptogenesis cannot generate sufficient asymmetry in the bulk of the
parameter space. The new paradigm is the quasi-thermal leptogenesis,
but this depends on the composition of the reheat plasma. If the
inflaton mainly decays to the top (s)quarks, the marginality between
gravitino production and leptogenesis will be improved (compared to
the case in full equilibrium).  As a result, the right amount of
baryon asymmetry can be generated for sufficiently heavy RH
(s)neutrinos. A more detailed and quantitative study is needed for a
better understanding of the role of decays and inverse decays in a
quasi-thermal leptogenesis.}

%%%%%%%%%%%%%%%%%%%%%%%%%%%%%%%%%%%%%%%%%%%%%%%%%%%%%%%%%%%%%%%%%%%
%%%%%%%%%%%%%%%%%%%%%%%%%%%%%%%%%%%%%%%%%%%%%%%%%%%%%%%%%%%%%%%%%%

\section{Thermalization after preheating}
\label{TAP}

So far we have focused on thermalization after the perturbative
inflaton decay. Although for final reheating the perturbative decay is
the most important, but we make some comments on the situation after a
non-perturbative decay of the inflaton also.  For detailed studies on
thermalization after preheating see Refs.~\cite{fk,mt,kp}.

Consider a simple chaotic inflation model as in
Appendix~\ref{perturbative} with the following potential
\beq \label{simple} V \sim {1 \over 2} m^2_{\phi}{\phi}^2 + h^2
{\phi}^2 {\chi}^2, 
\eeq 
where $\chi$ is another scalar field. Here we have considered only the
real parts of $\phi$ and $\chi$.  If $h > 10^{-6}$, the inflaton
oscillations decay to $\chi$ quanta via broad parametric
resonance~\cite{preheat2}. If $h > 10^{-4}$, resonant production
results in an extremely efficient transfer of energy from the
zero-mode condensate in a typical time scale $\sim 100 m^{-1}_{\phi}$
(which depends weakly on the coupling $h$)~\cite{preheat2}.

Resonant particle production and re-scatterings lead to the formation
of a plasma consisting of $\phi$ and $\chi$ quanta with typical
energies $\sim 10^{-1} \left(h m_{\phi} M_{\rm
P}\right)^{1/2}$~\cite{preheat2}. This plasma is in kinetic
equilibrium but full thermal equilibrium is established over a much
longer time scale than preheating~\cite{fk,mt}.

The occupation number of particles in the preheat plasma is $\gg 1$
(which is opposite to the situation after the perturbative
decay). This implies that the number density of particles is larger
than its value in full equilibrium, while the average energy of
particles is smaller than the equilibrium value.  It gives rise to
large effective masses for particles which, right after preheating, is
similar to their typical momenta~\cite{preheat2}.  Large occupation
numbers also lead to important quantum effects due to identical
particles and significant off-shell effects in the preheat plasma.
Because of all these, a field theoretical study of thermalization is
considerably more complicated in case of preheating. Due to the large
occupation numbers, one can consider the problem as thermalization of
classical fields at early stages~\cite{fk,mt,kp}.  In the course of
evolution towards full equilibrium, however, the occupation numbers
decrease. Therefore a proper (non-equilibrium) quantum field theory
treatment~\cite{bs} will be inevitably required at late stages when
occupation numbers are close to one.

Similar complications also arise when considering particle production
during thermalization. However let us make crude estimates based on
Eqs.~(\ref{gravquasi}) and~(\ref{baryonquasi}).  At the end of
preheating $\rho \sim 10^{-4} m^2_{\phi} M^2_{\rm P}$ and $T \sim
10^{-1} \left(h m_{\phi} M_{\rm P}\right)^{1/2}$. We therefore find
from Eq.~(\ref{a}) that, ${\cal A} \sim h^{-2} \gg 228.75$.
Eq.~(\ref{gravquasi}) then implies overproduction of gravitinos in the
preheat plasma unless, $T_{\rm max} \ll 10^{10}$ GeV. This results in
severe constraint on the models. For example, since $T_{\rm max} >
m_{\phi}$, it requires that $m_{\phi} \ll 10^{10}$ GeV which is a
disaster from the point of view of inflaton generating the density
perturbations. On the other hand, Eq.~(\ref{baryonquasi}), implies
that successful leptogenesis is now possible for $M \ll 10^9$~GeV. In
both the cases the situation is opposite to that after perturbative
decay where ${\cal A} \ll 1$.

We however caution the reader that these should be only taken as crude
estimates, since particle production from scatterings in the preheat
plasma is more involved. In a densed plasma the scattering processes
can be enhanced (for bosonic final states) or suppressed (for
fermionic final states). Large effective masses can also kinematically
suppress or shut-off some processes. We can nevertheless expect that
our analysis captures the main qualitative aspects of particle
production during thermalization after preheating.

%%%%%%%%%%%%%%%%%%%%%%%%%%%%%%%%%%%%%%%%%%%%%%%%%%%%%%%%%
%%%%%%%%%%%%%%%%%%%%%%%%%%%%%%%%%%%%%%%%%%%%%%%%%%%%%%%%%
\section{Conclusion}

In this paper we have presented a detailed account of the
thermalization after inflation in SUSY and we discussed various
implications. We have emphasized that the final stage of reheating is
the perturbative inflaton decay, even if the inflaton condensate
decays non-perturbatively. For a wide range of inflaton couplings
perturbative decay happens when the inflaton quanta dominate the
energy density and therefore generates entropy.

The most important result is the rate of thermalization is extremely
slow in SUSY. It is often wrongly assumed that the inflaton decay
products immediately thermalize if they have gauge interactions. Our
message is that this (although true in realistic models in non-SUSY
case) is {\it not} correct in SUSY.

In any SUSY extension of the SM there is a large number of flat
directions which are made up of squark, slepton and Higgs
fields. These flat directions acquire very large VEV during inflation
which spontaneously break gauge symmetries in the early Universe. This
induces very large masses to the gauge bosons (and gauginos) and
suppresses main reactions which lead to kinetic and chemical
equilibrium (i.e., the $2 \rightarrow 2$ and $2 \rightarrow 3$
scatterings with gauge boson exchange). As a result, the Universe
enters a long period of a quasi-thermal phase during which the
comoving number density and (average) energy of particles remain
constant.  This epoch lasts until the $2 \rightarrow 3$ scatterings
become efficient, at which point the number of particles increases and
full equilibrium is established. The main results are given in
Eqs.~(\ref{thermal1}) and~(\ref{thermal2}).

Slow thermalization substantially lowers the reheat temperature of the
Universe.  The reheat temperature is practically decoupled from the
inflaton decay and can be as low as ${\cal O}({\rm TeV})$, even for
large inflaton decay rates.  It varies in a typical range, $10^3~{\rm
GeV} \leq T_{\rm R} \leq 10^7$ GeV.  This is underlined in
Eq.~(\ref{rehtemp}) and demonstrated by the examples given in
Tables.~1,~2.

We studied particle production in a quasi-thermal phase. The general
results are given in
Eqs.~(\ref{kin}),~(\ref{hubble}),~(\ref{chidens}). Important aspect of
a quasi-thermal particle production is its dependence on the
composition of the reheat plasma (before complete thermalization). The
abundance of particles thus produced does not depend solely on the
maximum temperature in the quasi-thermal phase. We then specialized to
two cases of physical interest, namely gravitino production and
leptogenesis, with the results given in Eqs.~(\ref{gravquasi})
and~(\ref{baryonquasi}) respectively.

The most important cosmological consequence of our study is the
gravitino production. The Universe thermalizes at sufficiently low
reheat temperatures which satisfy the tightest BBN bounds on thermal
gravitino production.  Our central message is that there is a natural
resolution to the infamous Gravitino problem lies within a consistent
treatment of thermal history of the Universe within SUSY. We emphasize
that the built in solution offered by SUSY renders any exotic
modifications (such as late entropy release) unnecessary. Needless to
say, this has very important implications for inflationary model
building.

On the other hand, quasi-thermal leptogenesis is necessary to generate
sufficient baryon asymmetry since slow thermalization results in
$T_{\rm R} \ll 10^9$ GeV, for which thermal leptogenesis does not work
(unless in very special cases). As usual the question is the
marginality between leptogenesis and gravitino production. Depending
on the composition of the reheat plasma, this can be either relaxed or
tightened compared to the case in full equilibrium. If the inflaton
mainly decays into the top (s)quarks, it is possible to have
successful quasi-thermal leptogenesis while keeping gravitino
production under control. One can make further progress along these
lines through more quantitative studies which takes into account of
lepton-number violating interactions more carefully.

To conclude, supersymmetry dramatically modifies thermal history of
the Universe, most importantly, it provides a built in solution in the
form of flat directions which can naturally solve the gravitino
problem. This, so far, neglected fact can remove one of the most
serious obstacles for building consistent inflationary models in the
framework of SUSY and in string inspired theory.

%%%%%%%%%%%%%%%%%%%%%%%%%%%%%%%%%%%%%%%%%%%%%%%%%%%%%%%%%%%%%%%%%%%%%%%%%%%%%%
%%%%%%%%%%%%%%%%%%%%%%%%%%%%%%%%%%%%%%%%%%%%%%%%%%%%%%%%%%%%%%%%%%%%%%%%%%%%%%
\section{Acknowledgments}

The authors are thankful to Antonio Masiero for collaborating during
the initial stages of the work. Many of our discussions prompted into
various subsections and a major section which we have highlighted in
the text.

We would also like to thank Tirthabir Biswas, Wilfried Buchm\"uller,
Cliff Burgess, Robert Brandenberger, Jim Cline, Bhaskar Dutta, Andrew
Frey, Paolo Gondolo, Yuval Grossman, Richard Holman, Asko Jokinen, LeV
Kofman, Axel Krause, Alex Kusenko, Liam McAllister, Guy Moore, Takeo
Moroi, Marco Peloso, Dmitry Podolsky, Maxim Pospelov, Arvind Rajaraman
and Natalia Shuhmaher.  The work of R.A. is supported by the National
Sciences and Engineering Research Council of Canada. A.M. would like
to thank the hospitality of the ASPEN center for Physics, where
significant part of the work was carried out during the SUPERCOSMOLOGY
workshop. R.A. and A.M. would also like to thank the hospitality of
CITA, Perimeter Institute and McGill University, where quite a bit of
relevant physics were discussed while this work was being completed.

%%%%%%%%%%%%%%%%%%%%%%%%%%%%%%%%%%%%%%%%%%%%%%%%%%%%%%%%%%%%%%%%%%%%%%%%%%%%%%
%%%%%%%%%%%%%%%%%%%%%%%%%%%%%%%%%%%%%%%%%%%%%%%%%%%%%%%%%%%%%%%%%%%%%%%%%%%%%%
\section{Appendix}

\subsection{Last stage of inflaton decay} \label{perturbative}

Let us consider a simple model of chaotic inflation with a
SUSY~\footnote{A nice realization of chaotic inflation within
supergravity is through implementing a shift symmetry~\cite{kyy}. If
inflaton K\"ahler potential has the form $K = \left(\phi +
\phi^{\ast}\right)^2/M^2_{\rm P}$, instead of the minimal form $K =
\phi^{\ast} \phi/M^2_{\rm P}$, the scalar potential along the
imaginary part of $\phi$ remains flat even for Transplanckian field
values. Therefore it can play the role of inflaton in a chaotic
model. Note that a shift symmetry also ensures that the (positive)
Hubble induced corrections to the mass of flat directions vanishes at
the tree-level as the cross terms in Eq.~(\ref{cross}) disappear. It
is also possible to realize chaotic inflation for sub Planckian field
values in supergravity. For example, see the
multi-axions~\cite{Shamit} driven assisted inflation~\cite{Liddle}.}
superpotential
\beq \label{chasup}
W \supset {1 \over 2} m_{\phi} {\Phi} {\Phi} + h {\Phi} {\Psi} {\Psi},
\eeq
where ${\Phi}$ is the inflaton superfield comprising of the inflaton
$\phi$ and the inflatino ${\tilde \phi}$. It is coupled to another
superfield $\Psi$ whose bosonic and fermionic components are denoted
by $\chi$ and $\psi$ respectively. Here we choose $m_{\phi} = 10^{13}$
GeV, so that inflaton fluctuations generate the right amount of
density perturbations.  Eq.~(\ref{chasup}) results in the following
interaction terms in the scalar potential
\beq \label{infcoup}
V \supset h^2 {\phi}^2 {\chi}^2 + {1 \over \sqrt{2}} h m_{\phi} \phi 
{\chi}^2,    
\eeq
where we have considered the real parts of $\phi$ and $\chi$ fields. A
nice feature is that SUSY relates the couplings of the cubic $\phi
\chi^2$ and quartic $\phi^2 \chi^2$ interaction terms of the inflaton.
Note that the cubic term is required for complete decay of the
inflaton field.

At the end of inflation $\vert \phi \vert \sim {\cal O}(M_{\rm
P})$. Preheating occurs if $h > 10^{-6}$, in which case the $h^2
{\phi}^2 {\chi}^2$ term takes over and, for $h > 10^{-4}$, leads to an
explosive transfer of energy from the homogeneous condensate to $\chi$
quanta~\cite{preheat2}. Eventually, after re-scattering of $\chi$
quanta off the remaining condensate, a plasma is formed which consists
of the same number $\phi$ and $\chi$ quanta with typical energies $\gg
m_{\phi}$ which is in kinetic equilibrium~\cite{fk,mt,kp}.  This stage
completes over a rather short period of time $t \sim 100
m^{-1}_{\phi}$~\cite{preheat2}. Full thermal equilibrium takes much
longer to establish, but the temperature of the resulting thermal bath
will presumably be larger than $m_{\phi}$. This implies that the
inflaton (and inflatino) quanta remain in thermal equilibrium as long
as $T \gsim m_{\phi}$.

Once $T$ drops below $m_{\phi}$, due to Hubble expansion, the inflaton
quanta become non-relativistic.  The $h m_{\phi} \phi
{\chi}^2/\sqrt{2}$ term then takes over, leading to a perturbative
decay of the inflaton to (real and imaginary parts of) $\chi$, plus
the fermionic partner $\psi$, at a rate $\Gamma_{\rm d} = \left(h^2/8
\pi \right) m_{\phi}$. Note that regardless of how large $h$ is, this
stage of inflaton decay will be {\it perturbative}~\footnote{The
situation is similar to that in the decay of SUSY partners of SM
fields. These particles stay in thermal equilibrium at temperatures
above their mass. Once $T$ drops below their mass, they decay very
quickly, but {\it perturbatively} through gauge couplings of ${\cal
O}(1)$.}. The reason is that, unlike the initial condensate there is
no coherence among the decaying inflaton quanta at this stage.

The inflaton quanta dominate the energy density of the Universe at the
time of decay, and hence generate entropy, provided that
\beq \label{finpert}
\Gamma_{\rm d} \ll {m^2_{\phi} \over M_{\rm P}}.
\eeq
For $m_{\phi} = 10^{13}$ GeV, Eqs.~(\ref{infcoup}) and~(\ref{finpert})
result in $h < 10^{-2}$. This is much weaker than the condition $h
\leq 10^{-6}$, which is required for the inflaton decay to be
perturbative from the beginning.  Therefore in SUSY a last stage of
{\it perturbative} inflaton decay naturally follows preheating.
Let us consider a simple model of chaotic inflation with a
SUSY~\footnote{A nice realization of chaotic inflation within
supergravity is through implementing a shift symmetry~\cite{kyy}. If
inflaton K\"ahler potential has the form $K = \left(\phi +
\phi^{\ast}\right)^2/M^2_{\rm P}$, instead of the minimal form $K =
\phi^{\ast} \phi/M^2_{\rm P}$, the scalar potential along the
imaginary part of $\phi$ remains flat even for Transplanckian field
values. Therefore it can play the role of inflaton in a chaotic
model. Note that a shift symmetry also ensures that the (positive)
Hubble induced corrections to the mass of flat directions vanishes at
the tree-level as the cross terms in Eq.~(\ref{cross}) disappear. It
is also possible to realize chaotic inflation for sub Planckian field
values in supergravity. For example, see the
multi-axions~\cite{Shamit} driven assisted inflation~\cite{Liddle}.}
superpotential
\beq \label{chasup}
W \supset {1 \over 2} m_{\phi} {\Phi} {\Phi} + h {\Phi} {\Psi} {\Psi},
\eeq
where ${\Phi}$ is the inflaton superfield comprising of the inflaton
$\phi$ and the inflatino ${\tilde \phi}$. It is coupled to another
superfield $\Psi$ whose bosonic and fermionic components are denoted
by $\chi$ and $\psi$ respectively. Here we choose $m_{\phi} = 10^{13}$
GeV, so that inflaton fluctuations generate the right amount of
density perturbations.  Eq.~(\ref{chasup}) results in the following
interaction terms in the scalar potential
\beq \label{infcoup}
V \supset h^2 {\phi}^2 {\chi}^2 + {1 \over \sqrt{2}} h m_{\phi} \phi 
{\chi}^2,    
\eeq
where we have considered the real parts of $\phi$ and $\chi$ fields. A
nice feature is that SUSY relates the couplings of the cubic $\phi
\chi^2$ and quartic $\phi^2 \chi^2$ interaction terms of the inflaton.
Note that the cubic term is required for complete decay of the
inflaton field.

At the end of inflation $\vert \phi \vert \sim {\cal O}(M_{\rm
P})$. Preheating occurs if $h > 10^{-6}$, in which case the $h^2
{\phi}^2 {\chi}^2$ term takes over and, for $h > 10^{-4}$, leads to an
explosive transfer of energy from the homogeneous condensate to $\chi$
quanta~\cite{preheat2}. Eventually, after re-scattering of $\chi$
quanta off the remaining condensate, a plasma is formed which consists
of the same number $\phi$ and $\chi$ quanta with typical energies $\gg
m_{\phi}$ which is in kinetic equilibrium~\cite{fk,mt,kp}.  This stage
completes over a rather short period of time $t \sim 100
m^{-1}_{\phi}$~\cite{preheat2}. Full thermal equilibrium takes much
longer to establish, but the temperature of the resulting thermal bath
will presumably be larger than $m_{\phi}$. This implies that the
inflaton (and inflatino) quanta remain in thermal equilibrium as long
as $T \gsim m_{\phi}$.

Once $T$ drops below $m_{\phi}$, due to Hubble expansion, the inflaton
quanta become non-relativistic.  The $h m_{\phi} \phi
{\chi}^2/\sqrt{2}$ term then takes over, leading to a perturbative
decay of the inflaton to (real and imaginary parts of) $\chi$, plus
the fermionic partner $\psi$, at a rate $\Gamma_{\rm d} = \left(h^2/8
\pi \right) m_{\phi}$. Note that regardless of how large $h$ is, this
stage of inflaton decay will be {\it perturbative}~\footnote{The
situation is similar to that in the decay of SUSY partners of SM
fields. These particles stay in thermal equilibrium at temperatures
above their mass. Once $T$ drops below their mass, they decay very
quickly, but {\it perturbatively} through gauge couplings of ${\cal
O}(1)$.}. The reason is that, unlike the initial condensate there is
no coherence among the decaying inflaton quanta at this stage.

The inflaton quanta dominate the energy density of the Universe at the
time of decay, and hence generate entropy, provided that
\beq \label{finpert}
\Gamma_{\rm d} \ll {m^2_{\phi} \over M_{\rm P}}.
\eeq
For $m_{\phi} = 10^{13}$ GeV, Eqs.~(\ref{infcoup}) and~(\ref{finpert})
result in $h < 10^{-2}$. This is much weaker than the condition $h
\leq 10^{-6}$, which is required for the inflaton decay to be
perturbative from the beginning.  Therefore in SUSY a last stage of
{\it perturbative} inflaton decay naturally follows preheating.

%%%%%%%%%%%%%%%%%%%%%%%%%%%%%%%%%%%%%%%%%%%%%%%%%%%%%%%%%%%%%%%%%%%%%%%%%%%%%%
%%%%%%%%%%%%%%%%%%%%%%%%%%%%%%%%%%%%%%%%%%%%%%%%%%%%%%%%%%%%%%%%%%%%%%%%%%%%%%
\subsection{Gravitationally decaying inflaton}\label{hidden}

\footnote{This subsection is motivated by our discussion with Antonio
Masiero.}  As a first example, we consider a model of Ref.~\cite{rs}
in which the inflaton sector is gravitationally coupled to the MSSM
sector. The scalar potential in supergravity is given by~\cite{nilles}
\beq \label{scalpot}
V = e^G \left(G_i G^i
- {3 \over M^{2}_{\rm P}}\right) M^6_{\rm P},
\eeq
where $G$ is the K\"ahler function and in minimal supergravity is
defined as
\beq \label{kahler}
G = {{\chi_i} \chi^{*}_{i} \over M^{2}_{\rm P}} + 
\log \left ({{|W|}^{2} \over M^{6}_{\rm P}} \right ).
\eeq
The $\chi_i$ denote the scalar fields in the theory, and lower and
upper indices on $G$ denote its derivative with respect to $\chi_i$
and $\chi^*_i$ respectively. The inflaton sector superpotential
$W_{\phi}$ and the MSSM superpotential $W_{\rm MSSM}$ are given by
\beq \label{superpot1}
W_{\phi} = {1 \over 2} m_{\phi} \left(\Phi - M_{\rm P}\right)^2 ~ ~ ~ ~ ~ , 
~ ~ ~ ~ ~ W_{\rm MSSM} = 
y_{ijk} \Psi_{i} \Psi_{j} \Psi_{k}.
\eeq
Here $\Phi$ and $\Psi_{i}$ denote the inflaton and the MSSM chiral
superfields, respectively, and $y_{ijk}$ are the MSSM Yukawa
couplings. The minimum of inflaton potential is located at $\phi =
M_{\rm P}$, around which it takes the form $m^2_{\phi} \left(\phi -
M_{\rm P}\right)^2$. This model realizes new inflation in minimal
supergravity.  Obtaining density perturbations of the correct size
from quantum fluctuations of the inflaton requires that $m_{\phi} =
10^{13}$ GeV~\cite{rs}. Eq.~(\ref{scalpot}) leads to the following
term
\beq \label{scalcoup}
y_{ijk} {m_\phi \over M_{\rm P}} \phi^* \chi_i \chi_j \chi_k\,,
\eeq
in the scalar potential. If $m_{\phi}$ is much larger than the soft
SUSY breaking scalar masses, the partial width for inflaton decay to
three scalars is $\sim y^2 m^3_{\phi}/M^2_{\rm P}$. For $m_{\phi} =
10^{13}$ GeV this is always the case, particularly in models with weak
scale SUSY.

There is also a term
\beq \label{ferm}
e^{G/2} \left[G^{ij} - G^i G^j - G^k G^{ij}_{k}\right] {\bar \psi}_i 
{\psi}_j\,,
\eeq
in the Lagrangian~\cite{nilles} which describes the couplings of
fermionic partners of $\Psi$, denoted by $\chi$, in the two-component
notation
\beq \label{fermcoup}
y_{ijk} {1 \over M_{\rm P}} \phi \chi_i {\bar \psi}_j \psi_k\,.
\eeq
It results in a partial width for inflaton decay to two fermions and
one scalar which is same as that for decay to three scalars $\sim y^2
m^3_{\phi}/M^2_{\rm P}$. This implies that inflaton decay produces the
same number of particles and sparticles. Because of the large top
Yukawa coupling $y_t \approx 1$, the inflaton in this model mainly
decays to the top (s)quarks, LH bottom (s)quarks, Higgs $H_u$ and
Higgsino ${\widetilde H}_u$. The total inflaton decay rate is
therefore
\beq \label{decay1}
\Gamma_{\rm d} \sim 10^{-2} m^3_{\phi}/M^2_{\rm P}.
\eeq
Since the inflaton decays into three-body final states, and
hence the average energy of decay products is $\langle E \rangle
\approx m_{\phi}/3$, and hence $T_{\rm max} \approx m_{\phi}/9$. 
Eq.~(\ref{a}) then implies that 
\beq \label{a11}
{\cal A} \sim  \left({m_{\phi} \over M_{\rm P}}\right)^2\,.
\eeq
For $m_{\phi} = 10^{13}$ GeV, Eq.~(\ref{a11}) leads to ${\cal A} \sim
10^{-11}$. This implies that the reheat plasma is extremely dilute,
and hence substantially far from full equilibrium. For different
degrees of freedom we have:
\beq \label{a12} 
{\cal A}_i \sim y^2_i \left({m_{\phi} \over M_{\rm P}}\right)^2\,,
\eeq
where $y_i$ denotes the superpotential Yukawa coupling of the $i-$th
degree of freedom. Hence it is the largest for top (s)quarks, LH
bottom (s)quarks, $H_u$ and ${\widetilde H}_u$.

%For $m_{\phi} = 10^{13}$
%GeV, Eq.~(\ref{a11}) leads to ${\cal A} \sim 10^{-11}$. This implies an
%extremely dilute reheat plasma, and hence substantial deviation from
%full equilibrium.

%%%%%%%%%%%%%%%%%%%%%%%%%%%%%%%%%%%%%%%%%%%%%%%%%%%%%%%%%%%%%%%%%%%%%%%%%%%%%%
\subsection{Right-handed sneutrino as the inflaton}\label{sneutrino}

It is also possible that the inflaton is directly coupled to some of
the MSSM fields. This happens, for example, in the model of
Refs.~\cite{sninfl1,sninfl2} where one of the the RH sneutrinos
${\tilde N}$ plays the role of the inflaton. The relevant part of the
superpotential in this case reads:
\beq \label{superpot2}
W \supset {1 \over 2} M_N {\bf N} {\bf N} + {h}_{i} H_u 
{\bf N} L_i\,,
\eeq
where ${\bf N}$ is the multiplet containing the RH sneutrino $\tilde
N$ which plays the role of the inflaton (and its fermionic partner
$N$), and ${h}_{i}$ are the Yukawa couplings governing the inflaton
decay. With an appropriate choice of non-minimal K\"ahler function,
the scalar potential remains flat at large field values $\vert {\tilde
N} \vert > M_{\rm P}$, and this model realizes chaotic inflation in
supergravity~\cite{sninfl1,sninfl2} (see also~\cite{abdel}). Quantum
fluctuations of the sneutrino result in density perturbations of the
correct size, provided that $M_N = 10^{13}$ GeV.

The inflaton in this case mainly decays into the LH (s)leptons, $H_u$
and ${\widetilde H}_{u}$. Note that the same number of particles and
sparticles are produced in inflaton decay, so long as $M_N$ is much
larger than soft SUSY breaking masses (which is the case for weak
scale SUSY).  The total inflaton decay rate is then given by
\beq \label{decay2}
\Gamma_{\rm d} = {h^2 \over 4 \pi} M_N ~ ~ ; ~ ~ h \equiv 
\sqrt{\sum_{i}{{\vert h_i \vert}^2}}.  
\eeq
Since the inflaton decays into two-body final states, we have $\langle
E \rangle = m_{\phi}/2$ right after the decay completes, implying that
$T_{\rm max} \simeq m_{\phi}/6$.  Eq.~(\ref{a}) then results in
\beq \label{a21}
{\cal A} \sim 10^2 h^4 \left({M_{\rm P} \over M_N}\right)^2.
\eeq
For $M_N = 10^{13}$ GeV and $10^{-6} \leq h \leq 10^{-3}$ we find
$10^{-12} \leq {\cal A} \leq 1$. For the $i-$th (s)lepton doublet we
have
\beq \label{a22}
{\cal A}_i \sim 10^2 h^2_i h^2 \left({M_{\rm P} \over M_N}\right)^2.
\eeq
The (s)lepton singlets, (s)quarks, gauge fields and gauginos are not
produced in two-body decays of the inflaton. However they are
inevitably produced at higher orders of perturbation
theory~\cite{ad1}, therefore, they have much smaller but {\it
non-vanishing} values of ${\cal A}$.

%%%%%%%%%%%%%%%%%%%%%%%%%%%%%%%%%%%%%%%%%%%%%%%%%%%%%%%%%%%%%%%%%%%%%%%%%%%%%
%%%%%%%%%%%%%%%%%%%%%%%%%%%%%%%%%%%%%%%%%%%%%%%%%%%%%%%%%%%%%%%%%%%%%%%%%%%%%
\subsection{Flat directions and thermalization: additional considerations} 
\label{additional}

The main reason behind slow thermalization of the Universe in SUSY is
that the flat direction VEV, and hence the mass of gauge bosons,
remains large for a sufficiently long time.  In Section~\ref{TST-1} we
have assumed that the flat direction VEV is just redshifted by the
Hubble expansion while oscillating. Here we consider further details
of the flat direction dynamics and examine their importance.
%%%%%%%%%%%%%%%%%%%%%%%%%%%%%%%%%%%%%%%%%%%%%%%%%%%%%%%%%%%%%%%%%%%%%%%%%%%%%%
\begin{itemize}

\item{{\it Flat direction decay:}\\ In our analysis we have assumed
that flat directions do not decay until the Universe fully
thermalizes. The flat directions have gauge and Yukawa couplings,
generally denoted by $y$, to other fields which results in a decay
rate $\Gamma_{\varphi} = \left(y^2/4 \pi \right) m_0$. Note, however,
that the flat direction VEV induces a mass $y \vert \varphi \vert$ for
the decay products. The decay is therefore kinematically forbidden
until $y \vert \varphi \vert < m_0$.  On the other hand, the flat
direction decay to particles to which it is not directly coupled is
kinematically allowed at all times. However, such decays are mediated
by the fields which are coupled to the flat direction, and hence
suppressed by a factor $\sim \left(m_0/ y \vert \varphi \vert
\right)^2$ relative to the leading order decay.

In both cases it turns out that the flat direction decays at a time
when the expansion rate is given by:
\beq \label{flatdec}
H_{\varphi} < {1 \over 4 \pi} {m^3_0 \over {\vert \varphi \vert}^2}\,.
\eeq
The flat direction will not decay before the establishment of full
thermal equilibrium if $H_{\varphi} < H_{\rm thr}$ where, depending on
the case, $H_{\rm thr}$ is given by Eqs.~(\ref{thermal1})
and~(\ref{thermal2}). This is generically the case, in particular for
the examples of Table.~1.

One might also wonder whether the flat direction could promptly decay
via preheating since $y \vert \varphi \vert \gg m_0$ at the onset of
its oscillations~\cite{preheat2}.  However, SUSY breaking soft mass
terms in general lead to out-of-phase oscillations of the real and
imaginary parts of the flat direction~\cite{drt}. As shown
in~\cite{flatpreheat}, a tiny effect of this type is sufficient to
shut-off resonant decay of the flat direction. Similar effects will be
present if two or more flat directions with non-zero VEV oscillate
slightly out of phase.}

%%%%%%%%%%%%%%%%%%%%%%%%%%%%%%%%%%%%%%%%%%%%%%%%%%%%%%%%%%%%%%%%%%%%%%%%%%%%%%

\item{{\it Scatterings off the flat direction condensate:}\\ Energetic
particles in the reheat plasma scatter off the zero-mode quanta in the
flat direction condensate. In order for the flat direction to affect
thermalization, it should survive against evaporation by such
scatterings (at least) until the Universe fully
thermalizes. Scatterings which are mediated by gauge and gaugino
fields, similar to diagrams in Fig.~(1), play the main role
here. However, the center-of-mass energy in such scatterings is
$s^{1/2} \approx \left(4 E m_0 \right)^{1/2}$. It turns out that for
$H > H_{\rm thr}$ we typically have $4 E m_0 \ll \alpha {\vert \varphi
\vert}^2$.  The evaporation rate is therefore given by
\beq \label{flatevap}
\Gamma_{\rm eva} \sim 10 \alpha 
{n E m_0 \over {\vert \varphi \vert}^4},
\eeq
which is $< \Gamma_{\rm thr}$. This implies that the flat direction
evaporates only after the establishment of full equilibrium.

Note that the condensate contains zero-mode quanta with a number
density $n_{\varphi} = m_0 {\vert \varphi \vert}^2$. For sufficiently
large values of $\varphi_0$ that affect thermalization, $n_{\phi}$ is
much larger than the number density of particles in the reheat plasma
$n$. These zero-mode quanta can participate in the $2 \rightarrow 3$
scattering similar to those shown in Fig.~(2). Then one might wonder
whether scatterings off the flat direction condensate would result in
a thermalization rate larger than what we obtained in
Eq.~(\ref{chemical1}).

However emitting a gauge boson with a mass $\simeq g \vert \varphi
\vert$ requires that $4 E m_0 > \alpha {\vert \varphi \vert}^2$ (note
that $\left(4 E m_0 \right)^{1/2}$ is the center-of-mass energy). As
just mentioned, we have the opposite inequality for $H > H_{\rm thr}$.
This implies that the $2 \rightarrow 3$ scatterings off the condensate
do not have enough energy to produce on-shell gauge bosons. Inelastic
scatterings which happen at higher orders (mediated by off-shell gauge
and gaugino fields) can increase the number of particles but their
cross-section will be suppressed by (at least) a factor of $4 E m_0/\alpha 
{\vert
\varphi \vert}^2$ relative to those in Fig.~(2). This results in a
rate
\beq \label{flatthermal}
\Gamma^{\varphi}_{\rm thr} \sim 10 \alpha^2 {m^2_0 E \over 
{\vert \varphi \vert}^2}, 
\eeq
where we have used $n_{\varphi} = m_0 {\vert \varphi \vert}^2$. This
turns out to be subdominant to $\Gamma_{\rm thr}$, see
Eq.~(\ref{chemical1}), whenever the flat direction VEV is large enough
to affect thermalization.}

%%%%%%%%%%%%%%%%%%%%%%%%%%%%%%%%%%%%%%%%%%%%%%%%%%%%%%%%%%%%%%%%%%%%%%%%%%%%%%

\item{{\it Early oscillations of the flat direction:}\\ The flat
direction starts oscillating once the Hubble expansion rate drops
below its mass, which we have considered to be the soft breaking mass
$m_0$.

The thermal effects can trigger early oscillations of the flat
directions~\cite{ace,and}. This follows from a general consideration
that the fields which are coupled to a flat direction (once
kinematically accessible to a thermal bath) induce a plasma mass for
it. If the plasma mass exceeds the expansion rate of the Universe at
early times, it will trigger early oscillations of the flat
direction. When the reheat plasma is in full thermal equilibrium, this
indeed happens in many cases~\cite{ace,and}. Earlier oscillations of
the flat direction also imply earlier thermalization of the reheat
plasma, see Eqs.~(\ref{thermal1}) and~(\ref{thermal2}), and hence
higher reheat temperatures.

Before the establishment of full equilibrium, however, the reheat
plasma is dilute and the resulting plasma mass $\sim \left(n/E\right)^{1/2}$ 
is smaller by a factor of $\left({\cal
A}/228.75\right)^{1/4}$. Moreover, particles which are coupled to the
flat direction have a large mass, unless these couplings are very
small, and hence decay via gauge interactions almost instantly. Note
that the inverse decays are inefficient in a dilute plasma.  Therefore
the plasma masses are completely negligible and will not affect the
flat direction dynamics. }

\end{itemize}

%%%%%%%%%%%%%%%%%%%%%%%%%%%%%%%%%%%%%%%%%%%%%%%%%%%%%%%%%%%%%%%%%%%%%%%%%%%%%%
%%%%%%%%%%%%%%%%%%%%%%%%%%%%%%%%%%%%%%%%%%%%%%%%%%%%%%%%%%%%%%%%%%%%%%%%%%%%%%

\end{document}